\PassOptionsToPackage{table,xcdraw}{xcolor}
\documentclass[12pt,a4paper]{article}
\pdfoutput=1

\usepackage{geometry}
\usepackage[numbers,sort&compress]{natbib}
\usepackage{amsmath}
\usepackage{amssymb}
\usepackage[dvips]{graphicx}
\usepackage{pstricks}
\usepackage{bm}
\usepackage{pbox}
\usepackage{placeins}
\usepackage{graphicx}
\usepackage{caption}
\usepackage{subcaption}
\usepackage[T1]{fontenc}
\usepackage{footnote}
\usepackage{pdfpages}
\usepackage{hhline}
\usepackage{multirow}
\usepackage{enumitem}
\usepackage[nottoc,notlot,notlof]{tocbibind}
\usepackage[title,titletoc]{appendix}
\usepackage{dsfont}
\allowdisplaybreaks
\usepackage{cases}

\newcommand\scalemath[2]{\scalebox{#1}{\mbox{\ensuremath{\displaystyle #2}}}}

\definecolor{MyDarkBlue}{rgb}{0.1, 0.1, 0.8} 
\definecolor{MyLightBlue}{rgb}{0.22,0.51,0.9}
\definecolor{MyGreen}{rgb}{0.0, 0.5, 0.0}
\definecolor{BrickRed}{rgb}{0.8, 0.25, 0.33}
\usepackage[colorlinks=true, linkcolor=blue, citecolor=MyDarkBlue, urlcolor=MyLightBlue, bookmarksnumbered=true, bookmarksopen]{hyperref}
\hypersetup{colorlinks, citecolor=blue,linkcolor=red,  urlcolor=MyLightBlue}

\begin{document}
\vspace*{-0.2in}
\begin{flushright}
OSU-HEP-20-04
\end{flushright}
\vspace{0.5cm}

\begin{center}
{\Large\bf 
Combined explanations of $(g-2)_{\mu}$, $R_{D^{(\ast)}}$, $R_{K^{(\ast)}}$ anomalies \\ \vspace{0.2cm} in a two-loop radiative neutrino mass model
}\\
\end{center}

\vspace{0.5cm}
\renewcommand{\thefootnote}{\fnsymbol{footnote}}
\begin{center}
{\large
{}~\textbf{Shaikh Saad}\footnote{ E-mail: \textcolor{MyLightBlue}{shaikh.saad@okstate.edu}}
}
\vspace{0.5cm}

{\em Department of Physics, Oklahoma State University, Stillwater, OK 74078, USA }
\end{center}

\renewcommand{\thefootnote}{\arabic{footnote}}
\setcounter{footnote}{0}
\thispagestyle{empty}
\begin{abstract}
Motivated by the long-standing tension in the muon anomalous magnetic moment (AMM) and persistent observations of B-physics anomalies in  $R_{D^{(\ast)}}$ and $R_{K^{(\ast)}}$ ratios, we construct a simple  two-loop radiative neutrino mass model, and propose a combined  explanations of all these apparently disjoint phenomena within this  framework. Our proposed model consists of two scalar leptoquarks (LQs), a $SU(2)_L$ singlet $S_1\sim (\overline{3},1,1/3)$ and a $SU(2)_L$ triplet $S_3\sim (\overline{3},3,1/3)$ to accommodate  $R_{D^{(\ast)}}$ and $R_{K^{(\ast)}}$ anomalies, respectively. The muon receives chirality-enhanced contribution towards its $g-2$ due to the presence of $S_1$ LQ that accounts for the observed deviation from the Standard Model prediction. Furthermore, we introduce a $SU(2)_L$ singlet scalar diquark $\omega\sim (\overline{6},1,2/3)$, which is necessary to break lepton number and generate neutrino mass radiatively with the aid of $S_1$ and $S_3$ LQs. We perform a detailed phenomenological analysis of this set-up and demonstrate its viability by providing benchmark points  where a fit to the neutrino oscillation data together with proper explanations of the muon AMM puzzle and flavor anomalies are accomplished while simultaneously meeting all other flavor violation and collider bounds. 
\end{abstract}

\newpage
{\hypersetup{linkcolor=black}
\tableofcontents}
\setcounter{footnote}{0}
\clearpage
\section{Introduction}\label{SEC-01}
In the Standard Model (SM), contributions to the anomalous magnetic moment (AMM) of the muon $a_\mu$, arising from loop corrections \cite{Schwinger:1948iu} are calculated with excellent accuracy. On top of that since experiments determine this quantity to high precision, any deviation from the theory prediction directly points towards physics beyond the SM (BSM). In fact, there is a long-standing discrepancy between the theoretical computations \cite{Jegerlehner:2009ry, Keshavarzi:2018mgv, Davier:2019can} and its measured value \cite{Bennett:2006fi}, 
\begin{align}
\Delta a_{\mu}= a_\mu^{\text{exp}}-a^{\text{SM}}_\mu= (2.74\pm 0.73)\times 10^{-9},\label{amu}
\end{align}
corresponding to a $3.7\sigma$ anomaly. In the coming days, the Muon $g-2$ Collaboration \cite{Grange:2015fou}  at Fermilab is expected to announce their result, which further motivates our investigation of the possible NP explanation of this anomaly.

Over the last two decades, various mechanisms are proposed to account for this deviation. Among them the effects of scalar leptoquarks (LQs) on $a_\mu$ are studied extensively, see for example Refs. \cite{Djouadi:1989md, Davidson:1993qk, Couture:1995he, Cheung:2001ip} for single LQ solution to $(g-2)_\mu$. LQ extensions of the SM has gained a lot of attention recently, due to their ability in accommodating the persistent tensions observed in the lepton flavor
universality violating B meson decays, particularly in the $R_{K^{(\ast)}}$ and $R_{D^{(\ast)}}$ ratios (see for example Refs. \cite{Dorsner:2013tla, Sakaki:2013bfa, Duraisamy:2014sna,  Hiller:2014yaa, Buras:2014fpa,Gripaios:2014tna, Freytsis:2015qca, Pas:2015hca, Bauer:2015knc,Fajfer:2015ycq,  Deppisch:2016qqd, Li:2016vvp, Becirevic:2016yqi,Becirevic:2016oho, Sahoo:2016pet, Bhattacharya:2016mcc, Duraisamy:2016gsd, Barbieri:2016las,Crivellin:2017zlb, DAmico:2017mtc,Hiller:2017bzc, Becirevic:2017jtw, Cai:2017wry,Alok:2017sui, Sumensari:2017mud,Buttazzo:2017ixm,Crivellin:2017dsk, Guo:2017gxp,Aloni:2017ixa,Assad:2017iib, DiLuzio:2017vat,Calibbi:2017qbu,Chauhan:2017uil,Cline:2017aed,Sumensari:2017ovu, Biswas:2018jun,Muller:2018nwq,Blanke:2018sro, Schmaltz:2018nls,Azatov:2018knx, Sheng:2018vvm, Becirevic:2018afm, Hati:2018fzc, Azatov:2018kzb,Huang:2018nnq, Angelescu:2018tyl, DaRold:2018moy,Balaji:2018zna, Bansal:2018nwp,Mandal:2018kau,Iguro:2018vqb, Fornal:2018dqn, Kim:2018oih, deMedeirosVarzielas:2019lgb, Zhang:2019hth, Aydemir:2019ynb, deMedeirosVarzielas:2019okf, Cornella:2019hct,Datta:2019tuj, Popov:2019tyc, Bigaran:2019bqv, Hati:2019ufv, Coy:2019rfr, Balaji:2019kwe, Crivellin:2019dwb, 
Altmannshofer:2020axr, Cheung:2020sbq, Saad:2020ucl, Dev:2020qet} for both scalar and vector LQs explanations). These anomalies include flavor changing neutral current $b\to s$, as well as flavor changing charged current $b\to c$ transitions, which we briefly summarize below.    

Recent measurements have observed notable digressions from the SM predictions in the following two ratios associated with neutral current transition:
\begin{align}
R_K=\frac{\Gamma(\overline{B}\to \overline{K}\mu^+\mu^-)}{\Gamma(\overline{B}\to \overline{K}e^+e^-)},\;\;\; R_{K^*}=\frac{\Gamma(\overline{B}\to \overline{K}^{\ast}\mu^+\mu^-)}{\Gamma(\overline{B}\to \overline{K}^{\ast}e^+e^-)}.   
\end{align}
Theory predictions of these ratios are:
\begin{align}
&R^{\text{SM}}_K=1.0003\pm 0.0001\;\; \text{\cite{Bobeth:2007dw}},\;\;\;
R^{\text{SM}}_{K^*}=1.00\pm 0.01\;\;\text{\cite{Bordone:2016gaq}}.
\end{align}
On the contrary, the combined results of Run-1 data and Run-2 data of LHCb finds:
\begin{align}
&R^{\text{exp}}_K=
0.846^{+0.06+0.016}_{-0.054-0.014}, \;\;\tt{ 1.1\; GeV^2<q^2<6.0\; GeV^2}\;\;\text{\cite{Aaij:2019wad}},
\end{align}
for the $R_K$ ratio. Here the first uncertainty is statistical and the second uncertainty is systematic, and dilepton invariant mass squared is represented by $q^2$. This amounts to a tension of about $\gtrsim 2.5 \sigma$ between the theory and experiment. As for the $R_{K^*}$ ratio, the Belle collaboration finds the following values at low and high $q^2$ bins:
\begin{align}
&R^{\text{exp}}_{K^*}=
\begin{cases}
0.90^{+0.27}_{-0.21}\pm 0.10, \;\;\tt{ 0.1\; GeV^2<q^2<8.0\; GeV^2}\;\;\text{\cite{Abdesselam:2019wac}},\\
1.18^{+0.52}_{-0.32}\pm 0.10,  \;\;\tt{15\; GeV^2<q^2<19\; GeV^2} \;\;\text{\cite{Abdesselam:2019wac}}.
\end{cases}
\end{align}
Even though these values are in harmony with the SM, results from the LHCb show significant deviations compared to theory predictions, 
\begin{align}
&R^{\text{exp}}_{K^*}=
\begin{cases}
0.660^{+0.110}_{-0.070}\pm 0.024, \;\;\tt{0.045\; GeV^2<q^2<1.1\; GeV^2}\;\;\text{\cite{Aaij:2017vbb}},\\
0.685^{+0.113}_{-0.069}\pm 0.047,  \;\;\tt{1.1\; GeV^2<q^2<6.0\; GeV^2}\;\;\text{\cite{Aaij:2017vbb}}.
\end{cases}
\end{align}
These measured values in both low and high $q^2$ bins point towards $\gtrsim 2.5 \sigma$ deviation from SM values. Discrepancies observed in the $R_K$ and $R_{K^*}$ ratios have gained much curiosity in the theory community due to their trustable theory predictions, since hadronic uncertainties cancel out in these ratios.

Concerning the charged current transitions, experiments have observed noteworthy deviations from the SM values in the following two ratios: 
\begin{align}
&R_D=\frac{\Gamma(\overline{B}\to D\tau\nu)}{\Gamma(\overline{B}\to D\ell\nu)}, \;\;\;
R_{D^*}=\frac{\Gamma(\overline{B}\to D^{\ast}\tau\nu)}{\Gamma(\overline{B}\to D^{\ast}\ell\nu)}.
\end{align}
The corresponding SM predicted values of these quantities are, 
\begin{align}
&R^{\text{SM}}_D=0.299\pm 0.003\;\;\text{\cite{Na:2015kha, Aoki:2016frl}}, \;\;\;
R^{\text{SM}}_{D^*}=
0.258\pm 0.005\;\;\text{\cite{Bigi:2017jbd, Jaiswal:2017rve, Bernlochner:2017jka}}.
\end{align}
Persistent disagreement when compared to the SM predicted values in these ratios have been measured independently by several  different experiments. Deviations in the $B\to D\tau \nu$ transition are  observed by Babar \cite{Lees:2012xj, Lees:2013uzd} and Belle \cite{Huschle:2015rga, Sato:2016svk, Hirose:2016wfn, Abdesselam:2019dgh}, whereas discrepancies in the  $B\to D^{\ast}\tau \nu$ transition are measured by Babar \cite{Lees:2012xj, Lees:2013uzd}, Belle \cite{Huschle:2015rga, Sato:2016svk, Hirose:2016wfn, Abdesselam:2019dgh},  and LHCb \cite{Aaij:2015yra, Aaij:2017uff} collaborations. The combined world averages of these measurements amount to:
\begin{align}
&R^{\text{exp}}_D=
0.334\pm 0.031\;\;\text{\cite{Amhis:2016xyh, Abdesselam:2019dgh, Belle2019, Belle2019b}}, \;\;\;
R^{\text{exp}}_{D^*}=
0.297\pm 0.015\;\;\text{\cite{Amhis:2016xyh, Belle2019, Belle2019b}}.
\end{align}
Experimental results of $R_D$ and $R_{D^*}$ ratios indicate  a tension of about  $\gtrsim 3\sigma$ from the SM predictions. These observed significant deviations are also taken seriously in the particle physics community because the  corresponding SM calculations are reliable as these ratios are largely insensitive \cite{Bernlochner:2017jka} to hadronic uncertainties.

The outstanding tension of the muon AMM together with the large deviations measured in the lepton flavor universality violating  decays of the B mesons clearly indicate the existence of new physics beyond the SM. As already aforementioned, scalar leptoquarks are the prime candidates in resolving these observed  anomalies. However, a single scalar LQ cannot accommodate for three of these anomalies simultaneously. First, we identify the pair of LQs that can do our desired job. For a TeV scale LQ, a large enough contribution is required to account for $\Delta a_\mu$ data, which can be provided if both the left-handed and right-handed chiral couplings of the LQ are present \cite{Cheung:2001ip}.  This requirement is satisfied by only two scalar LQs, $S_1\sim (\overline{3},1,1/3)$ and $R_2\sim (3,2,7/6)$. It is interesting to realize that either of these two LQs can accommodate  anomalies in the $R_D$ and $R_{D^*}$ ratios at the tree-level (see for example Ref. \cite{Angelescu:2018tyl}). On the other hand,  $S_3\sim (\overline{3},3,1/3)$ is the only scalar LQ that can correctly incorporate $R_K$ and $R_{K^*}$ anomalies at the tree-level (see for example Ref. \cite{Angelescu:2018tyl}).

By following the above discussion, in this work, we postulate that the NP beyond the SM contains  $S_1$ and $S_3$ LQs. With these in hand, one must ask the obvious question: how to give mass to the neutrinos\footnote{Instead of $S_1$, if $R_2$ is used in association  with $S_3$, neutrino mass generation and reconciling B-physics anomalies are discussed in Refs. \cite{Popov:2019tyc, Saad:2020ucl}.}? The reason for this is, even though neutrinos remain massless in the SM, observations of neutrino oscillations are securely established by a number of  experiments \cite{Fukuda:1998mi, Ahmad:2002jz, Abe:2011sj, Adamson:2011qu, Abe:2011fz, An:2012eh, Ahn:2012nd}. Hence, any BSM construction is obliged to explain the origin of neutrinos masses and mixings. It gives rise to a more appealing scenario if the BSM states introduced to resolve these tensions also participate in neutrino mass generation mechanism\footnote{ 
See for example Refs. \cite{Pas:2015hca, Deppisch:2016qqd, Cai:2017wry, Guo:2017gxp, Hati:2018fzc, Datta:2019tuj, Popov:2019tyc, Saad:2020ucl, Dev:2020qet} that unify neutrino mass generation mechanism with B-physics anomalies.}. Since proper explanations of the above-mentioned anomalies demand TeV LQs, it is evident that the only natural choice to generate neutrino mass is via quantum corrections \cite{Cheng:1977ir, Zee:1980ai, Cheng:1980qt, Zee:1985id, Babu:1988ki, Babu:1988ig, Ma:2006km}. However, with just  $S_1$ and $S_3$ LQs added to the SM, neutrinos cannot get mass\footnote{Extension of the SM with $S_1$ and $S_3$ LQs was considered in Ref. \cite{Crivellin:2019dwb} without addressing the question of neutrino mass generation.  On the other hand, in Ref. \cite{Bigaran:2019bqv}, vectorlike-quarks $\sim (3,2,-5/6)$ was introduced in addition to  $S_1$ and $S_3$ LQs to give neutrinos non-zero masses.}. We must introduce one more BSM particle in the theory. One  obvious and simple choice is to extend the scalar sector by a color sextet diquark\footnote{Ref. \cite{Babu:2001ex} first proposed neutrino mass generation at the two-loop by introducing $S_1$ and $\omega$. This model is then analysed in more details and collider implications of these new colored states are studied in Ref. \cite{Kohda:2012sr}. Ref. \cite{Guo:2017gxp} considered the scenario of utilizing $S_3$ instead of $S_1$ in neutrino mass generation and to incorporate only $R_{K^{(*)}}$ anomaly. Furthermore, Ref. \cite{Datta:2019tuj} had the same particle content as that of Ref. \cite{Guo:2017gxp} and their work focused on explaining $R_{K^{(*)}}$ and $B\to K \pi$ anomalies. None of these frameworks can simultaneously explain the tensions in the $R_{D^{(*)}}$ and $a_\mu$, which we attempt to achieve in this work.} $\omega\sim (\overline{6},1,2/3)$, which is a singlet under the $SU(2)_L$. Addition of this scalar diquark (DQ) breaks the lepton number by two units, and Majonara mass for the neutrinos are then generated at the two-loop level, in which all three BSM particles run through the loop.

In a nutshell, we propose a framework in which the neutrino mass, the muon anomalous magnetic moment puzzle, and B-physics anomalies in the $R_{D^{(*)}}$, $R_{K^{(*)}}$ ratios have a common origin. We perform a comprehensive phenomenological analysis of this set-up and discuss the feasibility of interpreting these anomalies as well as explaining the neutrino oscillation data. 

In the next section (Sec. \ref{SEC-02}), we introduce the model and then discuss how to ameliorate these anomalies in Sec. \ref{SEC-03}. Relevant experimental constraints on the model parameters are detailed in Sec. \ref{SEC-04}. We present the results in Sec. \ref{SEC-05} and finally conclude in Sec. \ref{SEC-06}.

\section{The Set-up}\label{SEC-02}
\begin{figure}[th]
\centering
\includegraphics[scale=0.4]{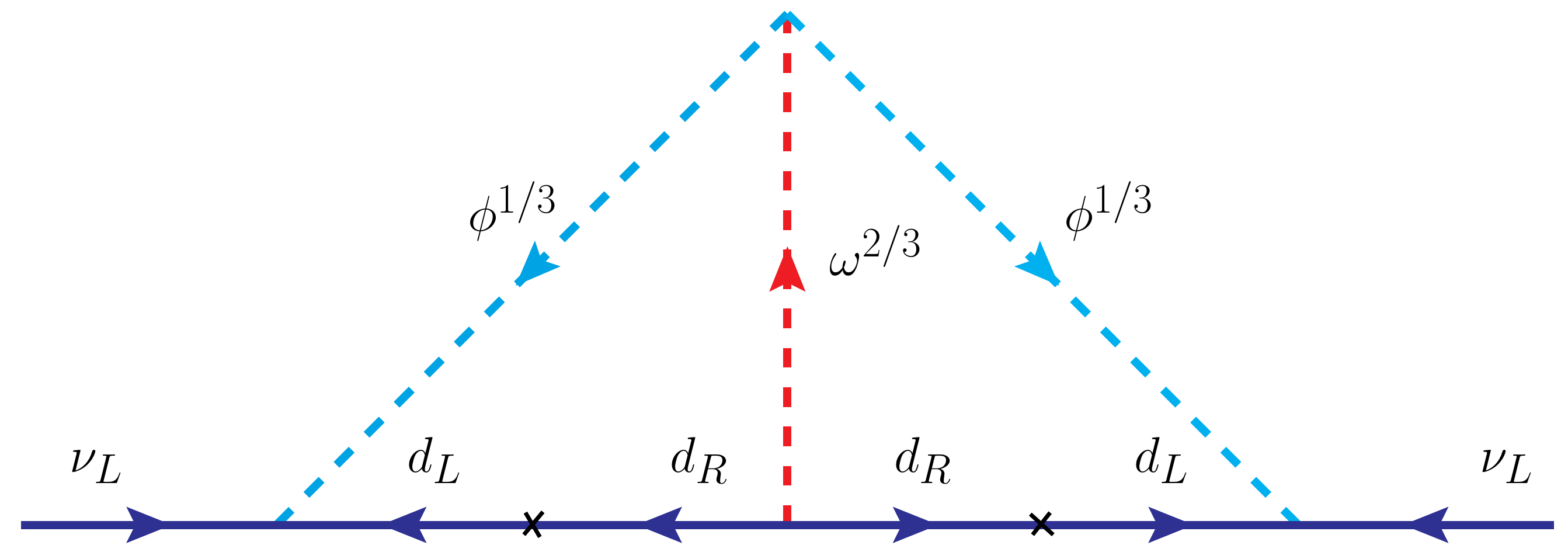}
\caption{Representative Feynman diagram for generating neutrino masses at the two-loop order. There are two separate diagrams, one  corresponding to $\phi=S^{1/3}_1$, and for the other $\phi=S^{1/3}_3$.} 
\label{nu-diag}
\end{figure}

Our proposed model consists of three BSM scalars, a $SU(2)_L$ singlet LQ $S_1\sim (\overline{3},1,1/3)$, a $SU(2)_L$ triplet LQ $S_3\sim (\overline{3},3,1/3)$, and a $SU(2)_L$ singlet DQ $\omega\sim (\overline{6},1,2/3)$. $S_1$ and $S_3$ LQs are introduced to accommodate the $R_{D^*}$ and $R_{K^*}$ flavor anomalies, respectively. The existence of $S_1$ LQ can account for the anomaly observed in the muon AMM $a_{\mu}$. Furthermore, both these LQs accompanied by the DQ $\omega$  participate in generating neutrino mass radiatively at the two-loop level, as shown in Fig. \ref{nu-diag}. As already aforementioned, existence of the DQ is required to break lepton number by two units, and provide mass to the neutrinos. Hence, in our model neutrinos are Majorana like fermions. The Yakawa couplings  associated to the LQs are given as follows \cite{Buchmuller:1986zs}: 
\begin{align}
\mathcal{L} \supset y_{ij}^L\; \overline{Q^c}_i i\sigma_2 S_1 L_j +y_{ij}^R\; \overline{u^c_R}_i  S_1 \ell_{R j} 
+y^S_{ij}\; \overline{Q^c}_i i\sigma_2 (\sigma^a S_3^a) L_j + \text{h.c.},
\label{LY}
\end{align}
as usual, here $Q$ and $L$ are the left-handed quark and lepton doublets of $SU(2)_L$, and $d_R$, $u_R$, and $\ell_R$ are the right-handed down-type quark, up-type quark, and charged lepton, respectively, which are all  singlets of $SU(2)_L$. Here $\sigma^a (a = 1,2,3$) are the Pauli matrices, and  $\{i,j\}$ are  flavor indices. Moreover, $S_3^a$ are the components of $S_3$ in the $SU(2)_L$ space. In the above Lagrangian, we have omitted the $S_{1,3}$ couplings to diquarks to ensure proton stability. The Yukawa couplings $y^L, y^R$ and $y^S$ are a priori arbitrary $3\times 3$ matrices in the flavor space.

To calculate the flavor observables, it is convenient to write the above Lagrangian in the charged fermion mass eigenbasis, for which we make the following transformations of the fermion fields:
\begin{align}
d_L \to d_L,\; u_L \to V^\dagger u_L,\;  \ell_L \to \ell_L,\; 
\nu_L \to U \nu_L \equiv \hat{\nu}_L.
\end{align}
Here  $U$ and $V$ are the Pontecorvo-Maki-Nakagawa-Sakata (PMNS) and Cabibbo-Kobayashi-Maskawa (CKM) matrices, respectively. Moreover, following the notation of Ref. \cite{Cai:2017wry} we have defined $\hat{\nu}$ that represents the neutrino weak-eigenstate.
With these, the Lagrangian takes the following form:
\begin{align}
&\mathcal{L}_{S_1} = -y^L_{ij} \overline{d^c_L}_i S_1^{1/3} \hat{\nu}_{Lj} + (V^* y^L)_{i j} \overline{u^c_L}_i S_1^{1/3} \ell_{Lj} + y^R_{ij}\overline{u^c_R}_i S_1^{1/3} \ell_{R j} + \text{h.c.}, \label{LagS1} \\
&\mathcal{L}_{S_3} = -y^S_{i j} \overline{d^c_L}_i S_3^{1/3} \hat{\nu}_{Lj} - \sqrt{2} y^S_{i j} \overline{d^c_L}_i S_3^{4/3} \ell_{Lj} + \sqrt{2} (V^* y^S)_{i j} \overline{u^c_L}_i S_3^{-2/3} \hat{\nu}_{Lj} \nonumber\\ 
&- (V^* y^S)_{i j} \overline{u^c_L}_i S_3^{1/3} \ell_{Lj} + \text{h.c.}, \label{LagS3}
\\ 
&\mathcal{L}_{\omega} = y^{\omega}_{ij} \overline{d^c_R}_i \omega d_{R j} + \text{h.c.} \label{LagOMG}
\end{align}
In the above set of Lagrangian terms, the Yukawa couplings of the DQ scalars are also summarized, which will be required for neutrino mass generation. Note that $y^{\omega}$ is a  $3\times 3$ symmetric matrix in the flavor space. 

Before proceeding any further, here we clarify few assumptions that we make. First, we assume all Yukawa couplings to be real for simplicity. The electroweak (EW) symmetry breaking will split the masses of the three component fields that belong to the triplet LQ. However, splittings among different components are highly constrained by EW precision measurements, this is why we chose them to be degenerate in mass. Subsequently, we ignore any mixing between the $S^{1/3}_1$ and $S^{1/3}_3$ states. These assumptions can be trivially guaranteed by appropriately choosing the corresponding quartic couplings in the scalar potential. Moreover, we denote the masses of the scalars by $M_1$, $M_3$, and $M_{DQ}$ for $S_1$, $S_3$, and $\omega$, respectively.   

In this given set-up, the neutrino mass generation occurs at the two-loop level via the diagrams as shown in Fig. \ref{nu-diag}. Note that there are two independent diagrams, one with $\phi=S^{1/3}_1$ and the other with $\phi=S^{1/3}_3$. These two-loop neutrino mass diagrams utilize the following cubic coupling terms in the scalar potential:
\begin{align}
V\supset \mu_1 S_1 S_1 \omega^* +    \mu_3 S_3 S_3 \omega^*  \supset \mu_1 S_1^{1/3} S_1^{1/3} \omega^{-2/3} +    \mu_3 S_3^{1/3} S_3^{1/3} \omega^{-2/3}.
\end{align}
With these, the neutrino mass formula has the following form \cite{Kohda:2012sr}: 
\begin{align}
\mathcal{M}^{\nu}_{ij} = 24 \mu_p \;y^p_{li}\; m^d_{ll}\; y^{\omega}_{lk}\; \mathcal{I}^p_{lk}\;  m^d_{kk}\; y^p_{kj}.\label{numatrix}
\end{align}
Here $m^d= diag\{m_d, m_s, m_b\}$ is the diagonal down-quark mass matrix. In this formula there are two terms, one for $p=1$ for which we have $\mu_p=\mu_1$, $y^p=y^L$, and the second for  $p=3$ that corresponds to $\mu_p=\mu_3$, $y^p=y^S$. Since the down-quark masses are very small compared to the LQ and DQ masses, the loop integrals have the following simple and generation independent form \cite{Babu:2002uu}:
\begin{align}
&\mathcal{I}^p_{lk}=  \frac{1}{256 \pi^4} \frac{1}{M^2_p} \mathcal{\overline{I}}\left[\frac{M^2_{DQ}}{M^2_p}\right], \label{I}\\
&\mathcal{\overline{I}}\left[r\right]=
\begin{cases}
\frac{\pi^2}{3};\;\;\;r\ll 1,\\
\frac{1}{r}\left( -1+\frac{\pi^2}{3}+(\log[r])^2 \right);\;\;\;r\gg 1.
\end{cases}
\end{align}
Here $M_p=M_1$ ($M_3$) for $p=1$ ($p=3$). Since the mass generation occurs at the two-loop level, TeV scale BSM states running in the loop naturally provide tiny masses to the neutrinos without requiring the Yukawa couplings  to be abnormally  small. In fact as we will show, Yukawa couplings of order $0.01 - 1$ are the prerequisite for concurrent explanations of B-physics anomalies, muon AMM, as well as  neutrino oscillation data.

\section{Resolving Anomalies}\label{SEC-03}
\subsection{\texorpdfstring{$(g-2)_{\mu}$}{muon AMM}}
\begin{figure}[th]
\centering
\includegraphics[scale=0.4]{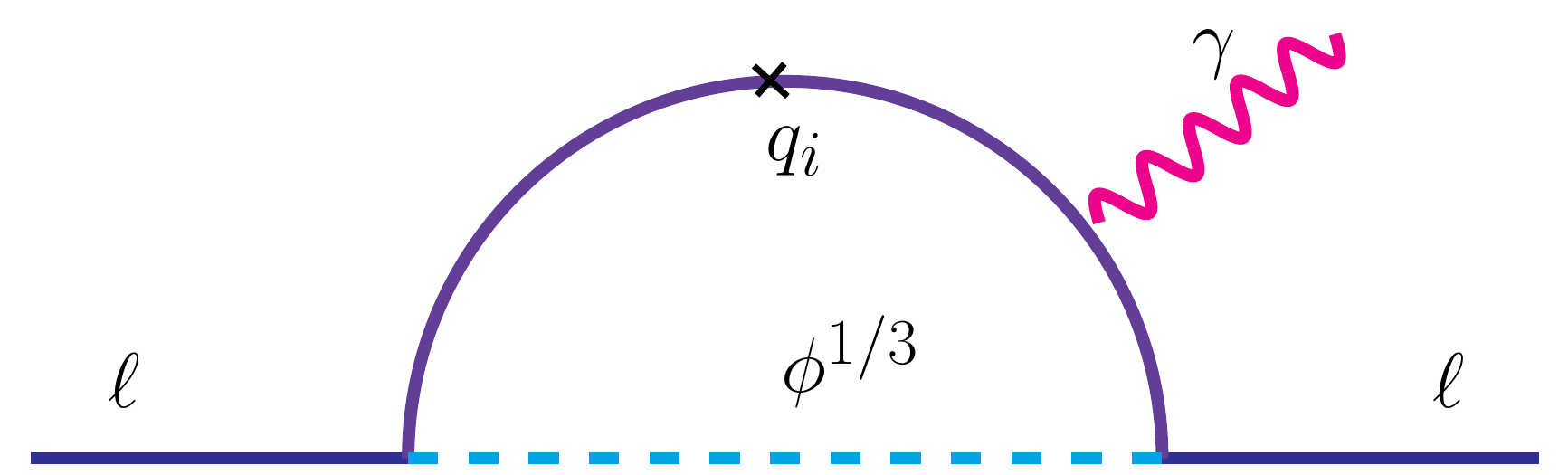}
\hspace{0.5cm}
\includegraphics[scale=0.4]{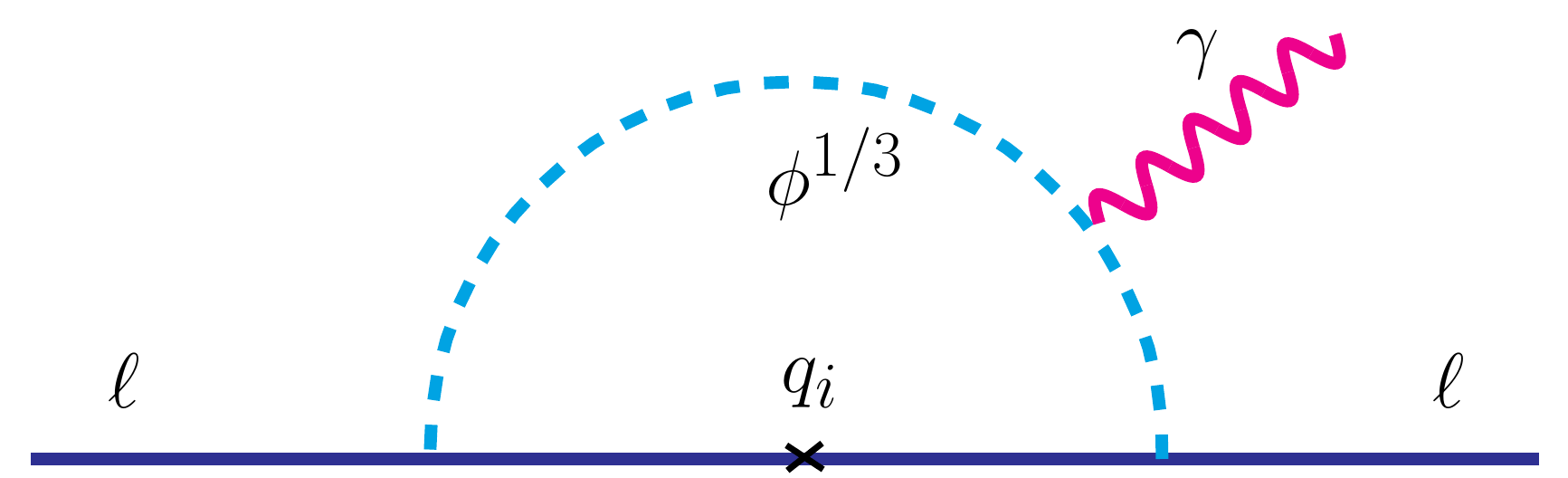}
\caption{Dominant contributions to the anomalous magnetic moments to charged leptons. } 
\label{g2MU}
\end{figure}

When considering LQ solutions to lepton AMMs, it is well known that  relevant contributions  can only be provided by a non-chiral LQs \cite{Cheung:2001ip} as discussed above. Among the only two non-chiral LQs $R_2$ and $S_1$, the latter is present in our set-up. Even though both $S_1$ and $S_3$ can in principle contribute to $a_{\ell}$ within our scenario, only effects coming from $S_1$  are important due to chiral enhancement. The dominant one-loop contributions to charged lepton AMM are presented in Fig. \ref{g2MU}.   

The effective Lagrangian from which $a_{\ell}$ is calculated  can be  written as \cite{Lavoura:2003xp, Cheung:2001ip, Dorsner:2016wpm}: 
\begin{align}
\mathcal{L}_{a_{\ell}} = e\overline{\ell} \left( \gamma^{\mu}A_{\mu} + \frac{a_{\ell}}{4 m_{\ell}}\sigma^{\mu\nu}F_{\mu\nu} \right) \ell, 
\end{align}
where the field strength tensor is defined as  $F_{\mu\nu}=\partial_{\mu}A_{\nu}-\partial_{\nu}A_{\mu}$, and the NP contribution to the AMM is calculated from $\Delta a_{\ell}=i\;m_{\ell}\left( \sigma_L +\sigma_R \right)$.  Here the contributions $\sigma_{L,R}$ can be computed from the effective Lagrangian that leads to $\ell\to \ell^{\prime}\gamma$, which is given below \cite{Lavoura:2003xp, Cheung:2001ip, Dorsner:2016wpm},
\begin{align}
\mathcal{L}_{\ell\to \ell^{\prime}\gamma}  =\frac{e}{2} \overline{\ell^{\prime}} i\sigma^{\mu\nu}F_{\mu\nu} \left(  \sigma_L^{\ell\ell^{\prime}}P_L + \sigma_R^{\ell\ell^{\prime}}P_R \right)\ell. \label{llp}  
\end{align}
To a very good approximation we find the corresponding contributions relevant to our study have the following expressions (by setting $\ell=\ell^{\prime}$):
\begin{align}
\sigma_{L,R}^{\ell\ell}=\frac{iN_c}{16\pi^2 M^2_1} \left( m_t V_{tb}y^L_{t\ell}y^R_{t\ell} \left[\frac{7}{6}+\frac{2}{3}\log[x_t]\right] \right),    
\end{align}
where we have assumed all couplings to be real. This leads to the following expression for the muon AMM arising dominantly from $S_1$ LQ:
\begin{align}
\Delta a_{\mu}\simeq-\frac{3}{8\pi^2} \frac{m_t m_{\mu}}{M^2_1} y^L_{32}y^R_{32}   \left[\frac{7}{6}+\frac{2}{3}\log[x_t]\right].
\end{align}
Here we have used the color factor $N_c=3$, $V_{tb}=1$, and $x_t=m^2_t/M_1^2$. Due to the top-quark mass insertion inside the loops as shown in Fig. \ref{g2MU}, the observed enhanced value of the muon magnetic moment can be naturally incorporated within this framework for a TeV scale leptoquark.

\subsection{\texorpdfstring{$R_K$ \text{and} $R_{K^*}$}{neutral current}}
\begin{figure}[th]
\centering
\includegraphics[scale=0.5]{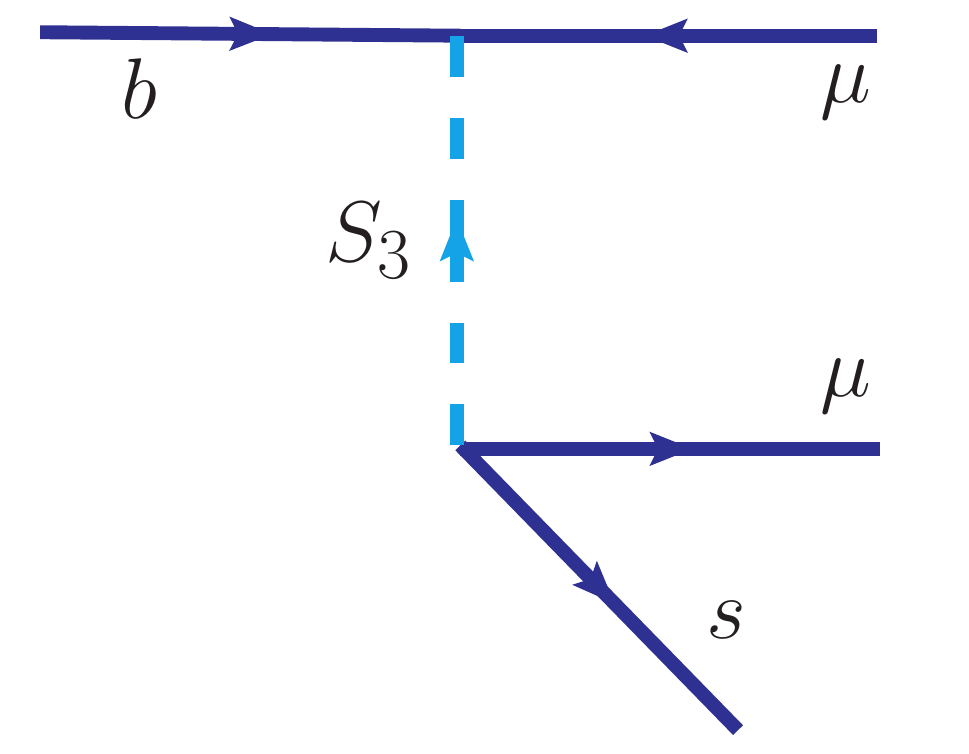}
\hspace{2cm}
\includegraphics[scale=0.5]{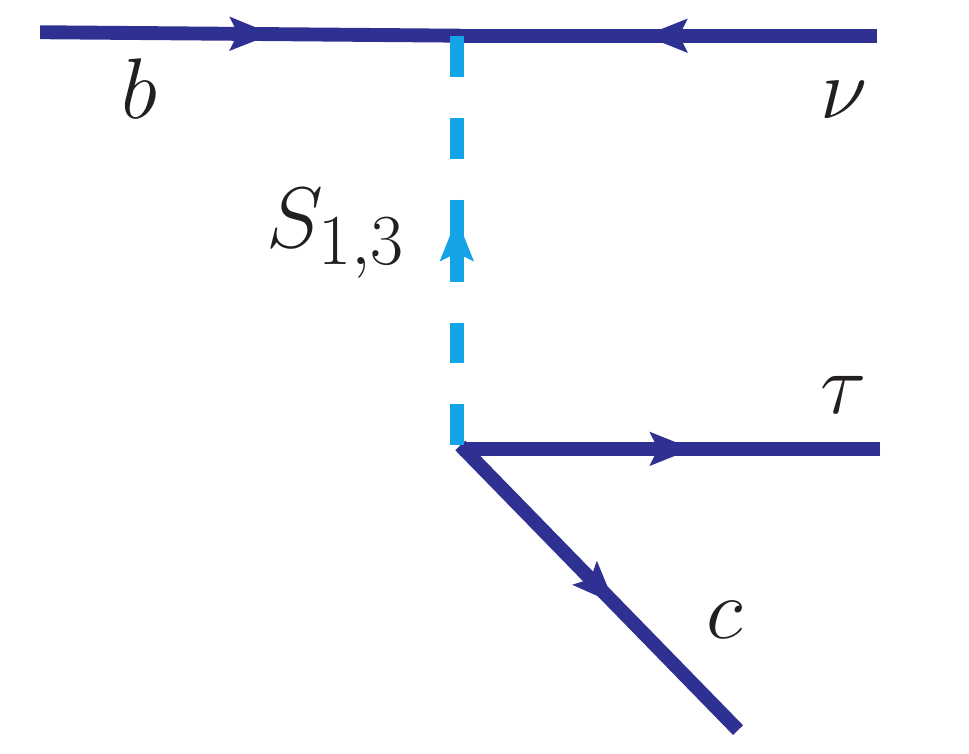}
\caption{Feynman diagrams representing  $b\to s\mu^-\mu^+$ (left) and  $b\to c\tau \overline{\nu}$ (right) transitions within our set-up.}
\label{RDRK}
\end{figure}

It is remarkable that $S_3$ is the only scalar LQ that can simultaneously account for $R_K < R_K^{\text{SM}}$  and $R_{K^*} < R_{K^*}^{\text{SM}}$ at tree-level. Processes of the form $B\to K^{(\ast)} \ell^+ \ell^{\prime -}$ can be described by the following effective Hamiltonian
\begin{align}
\mathcal{H}^{dd\ell\ell}_{eff}=-\frac{4 G_F}{\sqrt{2}} V_{tj}V^{\ast}_{ti}\left(\sum_{X=9,10}C_X^{ij,\ell\ell^{\prime}} \mathcal{O}_X^{ij,\ell\ell^{\prime}}\right)  +h.c., \label{Heff} 
\end{align}
where the effective operators are given by
\begin{align}
\mathcal{O}_9^{ij,\ell\ell^{\prime}}=\frac{\alpha}{4\pi}\left(\overline{d}_i\gamma^{\mu}P_Ld_j\right)\left(\overline{\ell}\gamma_{\mu}\ell^{\prime}\right),\;
\mathcal{O}_{10}^{ij,\ell\ell^{\prime}}=\frac{\alpha}{4\pi}\left(\overline{d}_i\gamma^{\mu}P_Ld_j\right)\left(\overline{\ell}\gamma_{\mu}\gamma_5\ell^{\prime}\right).
\end{align}
After integrating out the heavy leptoquark and combining the Yukawa part of the Lagrangian associated to $S_3$ as given in Eq. \eqref{LagS3} with the above effective Hamiltonian lead to the following purely vector Wilson coefficients at the LQ mass scale: 
\begin{align}
C^{\ell\ell^{\prime}}_{9}=-C^{\ell\ell^{\prime}}_{10}=\frac{v^{2}}{V_{t b} V_{t s}^{\ast}}\frac{\pi}{\alpha_{em}} \frac{y^S_{b\ell^{\prime}} \left(y^S_{s \ell} \right)^{*}}{M_{3}^{2}}.\label{RK1}
\end{align}
By assuming the NP coupling to electrons is negligible (leading to $\ell=\ell^{\prime}=\mu$), the observed values of the $R_K$ and $R_{K^*}$ ratios then can be explained with $C_{9,10}^{22} < 0$. Wilson coefficients of this type are generated within our framework by the $S_3$ LQ couplings to muons over the electrons as depicted in Fig. \ref{RDRK} (left diagram). 

In addition to $R_K$ and $R_{K^*}$ ratios, 
discrepancies are also founds in several other observables related to neutral current processes. For example the most significant departure has been found in the angular observable $P'_5$ in the $B\to K^*\mu\mu$  decay \cite{Aaij:2015oid, Khachatryan:2015isa, Aaboud:2018krd}. Another notable disagreement is in the  combined fit to the anomalous $b\to s$ data in operators contributing to $b\to s\mu\mu$ \cite{DAmico:2017mtc, Geng:2017svp, Capdevila:2017bsm, Altmannshofer:2017yso, Ciuchini:2017mik, Hiller:2017bzc, Aebischer:2019mlg, Ciuchini:2019usw, Kowalska:2019ley, Alok:2019ufo}.  

An excellent fit to the $R_{K^{(*)}}$ flavor ratios as well as the above-mentioned several other discrepancies associated with the neutral current transitions  can be found with $C^{22}_{9}=-C^{22}_{10}=-0.53$ \cite{Aebischer:2019mlg}, and the allowed range of values of these coefficients are $\left[-0.61,-0.45\right]$ ($1\sigma$ confidence level) and $\left[-0.69,-0.37\right]$ ($2\sigma$ confidence level).

\subsection{\texorpdfstring{$R_D$ \text{and} $R_{D^*}$}{charged current}}
As for the charged current process $b\to c\tau \overline{\nu}$ that is responsible for B meson decays $B\to D\tau\nu$ and $B\to D^{\ast}\tau\nu$ get contributions from both $S_1$ and $S_3$ LQs at the tree-level. Feynman diagrams that lead to such processes are shown in Fig. \ref{RDRK} (right diagram). Processes of these types can be described by considering the following effective Hamiltonian: \begin{align}
\begin{aligned}
\mathcal{H}^{du\ell\nu}_{\mathrm{eff}}=\frac{4 G_{F}}{\sqrt{2}} V_{cb} \left[C^{fi}_V\left(\overline{\ell}_{L} \gamma^{\mu} \nu_{Li}\right)\left(\overline{c}_{L} \gamma_{\mu} b_{L}\right)\right.&+C^{fi}_{S}\left(\overline{\ell}_{Rf} \nu_{Lj}\right)\left(\overline{c}_{R} b_{L}\right) 
\\
&\left.+C^{fi}_{T}\left(\overline{\ell}_{Rf} \sigma^{\mu \nu} \nu_{Li}\right)\left(\overline{c}_{R} \sigma_{\mu \nu} b_{L}\right)\right]+\mathrm{h.c.},
\end{aligned}
\end{align}  
where in the SM $C^{\text{SM}}_V=1$. In the above effective Hamiltonian, both $S_1$ and $S_3$ contribute to the vector Wilson coefficient, whereas only $S_1$ participates in the scalar and tensor Wilson coefficients, which at the LQ mass scale have the following forms:
\begin{align}
&C^{fi}_S=-4  C^{fi}_T= -\frac{v^2}{4V_{cb}} \frac{y^L_{bi}\left(y^R\right)^*_{cf}}{M^2_1},\label{RD1}
\\
&C^{fi}_V= \frac{v^2}{4V_{cb}} \left[    \frac{y^L_{bi}\left(V^*y^L\right)^*_{cf}}{M^2_1} -
\frac{y^S_{bi}\left(V^*y^S\right)^*_{cf}}{M^2_3} \right]. \label{RD2}
\end{align}
We will focus on scenarios where dominant coefficients are the ones with $i=3$, which corresponds to lepton flavor conservation \cite{Sakaki:2013bfa, Freytsis:2015qca, Bhattacharya:2015ida, Bardhan:2016uhr, Bhattacharya:2016zcw, Choudhury:2016ulr, Cai:2017wry}.    Then the expressions of the $R_D$ and $R_{D^*}$ ratios are given by \cite{Blanke:2018yud}:
\begin{align}
R_D\simeq &R^{\text{SM}}_D\left( \left| 1+C^{33}_V\right|^2 +1.54 Re\left[ (1+C^{33}_V)(C^{33}_S)^*\right] +1.09 \left|  C^{33}_S\right|^2 \right. 
\nonumber \\& \left. \hspace{90pt}
+1.04 Re\left[ (1+C^{33}_V)(C^{33}_T)^* \right] +0.75 \left| C^{33}_T\right|^2 \right),    
\\
R_{D^*}\simeq &R^{\text{SM}}_{D^*}\left( \left| 1+C^{33}_V\right|^2 -0.13 Re\left[ (1+C^{33}_V)(C^{33}_S)^*\right] +0.05 \left|  C^{33}_S\right|^2 \right. 
\nonumber \\& \left. \hspace{90pt}
-5.0 Re\left[ (1+C^{33}_V)(C^{33}_T)^* \right] +16.27 \left| C^{33}_T\right|^2 \right). 
\end{align}
In these formulas, the Wilson coefficients are given at the low scale $\mu=m_b$. 

In addition to $R_{D^{(*)}}$, there are a number of observables associated to the charged current processes that indicate disagreements to some extent when compared to the SM values. Such as the ratio $R_{J/\psi}$ of the tauonic mode to the muonic mode for $B\to J/\psi \ell \nu$ \cite{Aaij:2017tyk}, the longitudinal polarization of the $D^*$ denoted by $f^{D^*}_L$ \cite{Abdesselam:2019wbt}, and polarization asymmetry in the longitudinal direction of the tau in the $D^*$ mode denoted by  $\mathcal{P}^*_\tau$ \cite{Hirose:2016wfn}. These observables have comparatively large error bars, hence we focus only on explaining $R_{D^{(*)}}$.

\subsection{Synopsis}
Here we discuss the textures of the Yukawa coupling matrices required for a combined explanations of the aforementioned phenomena that we want to achieve in this work.  First note that three Yukawa coupling matrices, $y^{L,S,\omega}$ enter in the neutrino mass formula given in Eq. \eqref{numatrix}. Among them, $y^L$ participates in explaining both the muon AMM and $R_{D^{(*)}}$ anomalies, whereas $y^S$  takes part in incorporating  $R_{K^{(*)}}$ and $R_{D^{(*)}}$ ratios. The only Yukawa coupling matrix, $y^R$ that does not contribute to neutrino masses, however plays significant role in resolving tensions in $a_\mu$ and $R_{D^{(*)}}$.

As for the neutrinos, two mass squared differences and three mixing angles have been measured with great accuracy. Even though the hierarchical pattern, whether normal ordering ($m_3>m_2>m_1$) or inverted ordering ($m_2>m_1>m_3$) is not yet known, inverted ordering is less favored by the data. Hence in this work, we stick to normal ordering for the neutrinos. In addition to masses and mixings, if neutrinos are Majorana fermions, then there are three more physical quantities exists in the neutrino sector. One of them is the Dirac phase, and rest two are Majorana phases. Dirac phase, which has not been measured yet directly, currently has large uncertainty associated to it \cite{Abe:2019vii}, furthermore we have no clue about the range of the Majorana phases. In this work, we take all parameters to be real, and do not focus on predicting these phases. An overview of the most recent global fit \cite{Esteban:2018azc} to the neutrino oscillation data are given as follows:
\begin{align}
&\Delta m_{21}^2\;(10^{-5} eV^2)\;= 7.39_{-0.20}^{+0.21};\;\;\;
\Delta m_{31}^2\; (10^{-3} eV^2)\;= 2.523_{-0.030}^{+0.032};
\\
&\sin^2\theta_{12}= 0.31_{-0.012}^{+0.013}; \;\;\;
\sin^2\theta_{13}=  0.02241_{-0.00065}^{+0.00066}; \;\;\;   
\sin^2\theta_{23}=  0.558_{-0.033}^{+0.020}. 
\end{align}

Following the aforementioned discussions on reconciling these anomalies along with neutrino oscillation data, we adopt the following  form of the Yukawa coupling matrices:
\begin{align}
y^R=\left(\begin{array}{lll}
0 & 0 & 0 \\
0 & 0 & \red{\ast} \\
0 & \blue{\ast} & 0
\end{array}\right),\;
y^L = \left(\begin{array}{lll}
0 & 0 & 0 \\
0 & * & \red{\ast} \\
0 & \blue{\ast} & \red{\ast}
\end{array}\right),\;
y^S = \left(\begin{array}{lll}
0 & 0 & 0 \\
0 & \green{\ast} & \red{\ast} \\
* & \green{\ast} & \red{\ast}
\end{array}\right),
\;
y^{\omega} = \left(\begin{array}{lll}
0 & 0 & 0 \\
0 & * & * \\
0 & * & *
\end{array}\right). \label{matrix}
\end{align}
The entries in blue plays role in $\Delta a_{\mu}$, the entries in green enters into the $R_{K^{(*)}}$ expressions, and the couplings in red contribute to $R_{D^{(*)}}$ ratios. Additionally the entries in black are introduced to get consistent fit to the neutrino masses and mixing angles. A few comments are in order regarding the choice of the above Yukawa couplings matrices. Since explanations to the B meson decay anomalies demand the existence of most of the entries in the lower $2\times 2$ blocks, it is a natural choice to populate $y^{\omega}$ matrix in the same lower  $2\times 2$ block. We intentionally do not introduce any couplings with the first generation of quarks in the above matrices, since these couplings are severely constrained by many different experiments. However, it is easily understood that  filling out all entries in the lower $2\times 2$ block  are not sufficient to give a realistic fit to the neutrino data. Introducing $y^{\omega}_{21}$ or $y^{\omega}_{31}$ term does not change the above conclusion either.  This leaves us with four different minimal options, considering one non-vanishing term from the set $\{y^S_{21}, y^S_{31}, y^L_{21}, y^L_{31}\}$. Instead of exploring all such possibilities, we fix  $y^S_{31}\neq 0$ for the rest of the analysis. In the next section, we elucidate the experimental constraints on the aforementioned non-zero Yukawa couplings.

Before closing this section, here we briefly discuss the loop corrections and running of the Wilson coefficients.  The QCD corrections to the matching on 2-quark-2-lepton operators mediating semileptonic B decays have been recently computed in Ref. \cite{Aebischer:2018acj}. This correction leads to a shift of the Wilson coefficients Eqs. \eqref{RK1}, \eqref{RD1}, and \eqref{RD2} that are,
\begin{align}
&C_S\to C_S \left( 1+\frac{2\alpha_s}{\pi} \right), \\ 
&C_T\to C_T \left( 1+\frac{8\alpha_s}{2\pi} +\frac{4\alpha_s}{3\pi}\log\left[\frac{\mu^2}{M^2_{LQ}} \right] \right),\\
&C_V\to C_V \left( 1+\frac{17\alpha_s}{6\pi} +\frac{\alpha_s}{\pi}\log\left[\frac{\mu^2}{M^2_{LQ}} \right] \right).
\end{align}
These QCD corrections enhance the contributions to about $10\%$ \cite{Aebischer:2018acj} which definitely favor towards the explanations of B meson decay anomalies. Furthermore, to evaluate the above-mentioned flavor ratios, we run these operators to the bottom-quark mass scale at which the relevant form factors are calculated. We use the {\tt Flavio} package \cite{Straub:2018kue} to do this running (see also Ref. \cite{Gonzalez-Alonso:2017iyc}) and find the following relations between the two different scales:
\begin{align}
&C_S(\mu=m_b)= 1.646\;C_S(\mu=M_{LQ}),\\
&C_T(\mu=m_b)= 0.863\;C_T(\mu=M_{LQ}),\\
&C_V(\mu=m_b)= 1.0\;C_V(\mu=M_{LQ}).
\end{align}
In this calculation, we have fixed $M_{LQ}=1200$ GeV, and for the bottom-quark mass $m_b= 4.18$ GeV is used. The relation between the scalar and the tensor Wilson coefficients also gets modified at the low scale, which we find to be $C_S(\mu=m_b)= -7.63\; C_T(\mu=m_b)$.

\section{Correlated Observables}\label{SEC-04}
In the previous sections, we have discussed the NP contributions to the muon AMM and $R_{K^*}$, $R_{D^*}$ flavor ratios, and the  neutrino mass generation mechanism is introduced in Sec. \ref{SEC-02}. Accommodating these significant deviations from the theory predictions lead the way to various flavor violating processes that are severely constrained by experimental data. In this section, we consider all such relevant processes and discuss the associated constraints on the model parameters.    

\subsection{\texorpdfstring{$\ell\to \ell^{\prime}\gamma$}{l to l gamma} Processes}
The effective Lagrangian leading to radiative decays of the charged leptons $\ell\to \ell^{\prime}\gamma$ is given in Eq. \eqref{llp}. Although both LQs mediate these dangerous processes, $S_1$ mediated $\tau\to \mu \gamma$ receives chirality-enhanced effect from top-quark, which is the strongest constraint within our model. The branching ratios associated to these process are calculated by the following formula \cite{Lavoura:2003xp}:
\begin{align}
Br(\ell\to \ell^{\prime}\gamma) = \frac{\tau_{\ell}\;\alpha\; m^3_{\ell}}{4}  \left( |\sigma^{\ell\ell^{\prime}}_L|^2 + |\sigma^{\ell\ell^{\prime}}_R|^2 \right),
\end{align}
where $\tau_{\ell}$ is the lifetime of the initial state lepton and  we derive the following expressions of these $\sigma_{L,R}$  originating from $S_1$ and $S_3$ LQs \cite{Lavoura:2003xp, Dorsner:2016wpm}: 
\begin{align}
&\sigma_{L,S_1}^{if}=\frac{iN_c}{16\pi^2 M^2_1} \left\{ (Vy^L)^*_{qf}(Vy^L)_{qi}m_f\frac{-1}{12} +
(y^R)^*_{qf}(Vy^L)_{qi}m_q
\left( \frac{7}{6}+\frac{2}{3}\log(x_q) \right) \right\}, 
\\
&\sigma_{R,S_1}^{if}=\frac{iN_c}{16\pi^2 M^2_1} \left\{ (Vy^L)^*_{qf}(Vy^L)_{qi}m_i\frac{-1}{12} +
(Vy^L)^*_{qf}y^R_{qi}m_q
\left( \frac{7}{6}+\frac{2}{3}\log(x_q) \right) \right\}, 
\\
&\sigma_{L,S_3}^{if}=\frac{iN_c}{16\pi^2 M^2_3}m_i \left\{ 
(Vy^S)^*_{qf}(Vy^S)_{qi}\frac{-1}{12}+(y^S)^*_{qf}y^S_{qi}\frac{1}{3}
\right\}, 
\\
&\sigma_{R,S_3}^{if}=\frac{iN_c}{16\pi^2 M^2_3}m_f \left\{ 
(Vy^S)^*_{qf}(Vy^S)_{qi}\frac{-1}{12}+(y^S)^*_{qf}y^S_{qi}\frac{1}{3}
\right\}.
\end{align}
Here as before $x_q=m^2_q/M^2_{LQ}$. However, for $q\neq t$, the replacement of $m_q\to \mu_{LQ}$ inside the $\log$-function needs to be made for consistency, see Ref. \cite{Crivellin:2019dwb} for details. In the above formulas terms proportional to $y^R(y^R)^*$ are not shown, since they vanish for our choice of the Yukawa textures. In the following we summarize the current experimental limits on these processes \cite{ TheMEG:2016wtm, Aubert:2009ag}:
\begin{align}
&Br\left(\mu\to e\gamma\right)<4.2\times 10^{-13},\\ 
&Br\left(\tau\to e\gamma\right)<3.3\times 10^{-8}, \\
&Br\left(\tau\to \mu\gamma\right)<4.4\times 10^{-8}.
\end{align}

\subsection{\texorpdfstring{$\ell\to \ell^{\prime}\ell^{\prime}\ell^{\prime\prime}$}{l to 3l} Processes}
The interaction terms that lead to $\ell\to \ell^{\prime}\gamma$ also generate the rare lepton flavor violating decays $\ell\to \ell^{\prime}\ell^{\prime}\ell^{\prime\prime}$. LQs present in our set-up induce these processes at the one-loop level. Decays of these types proceed via penguin-diagrams with $Z$ and $\gamma$ exchanges, and via box-diagrams with quarks and LQs inside the loops. The corresponding box-diagram contributions are always somewhat smaller than the penguin-diagrams, hence we omit those terms. Branching ratios of such decay channels can be written as  \cite{Arganda:2005ji, Benbrik:2010cf, deBoer:2015boa, Abada:2014kba, Mandal:2019gff}:
\begin{align}
Br(\ell_l^- \to (3\ell_n)^-)=
&\frac{\alpha_e^2m_{\ell_i}^5}{32\pi\Gamma_{\ell_i}}\;
\bigg\{ |T_{1L}|^2+|T_{1R}|^2+\left(|T_{2L}|^2+|T_{2R}|^2\right)\left(\frac{16}{3}\ln\frac{m_{\ell_i}}{m_{\ell_n}}-\frac{22}{3}\right)\nonumber\\
&-4\,\mathrm{Re}[T_{1L}T_{2R}^*+T_{2L}T_{1R}^*]
+\frac13\left(2\left(|Z_L g_{Ll}|^2+|Z_R g_{Rl}|^2\right)+|Z_L g_{Rl}|^2+|Z_R g_{Ll}|^2\right)\nonumber\\
&+\frac23\,\mathrm{Re}[2\, (T_{1L}Z_L^*g_{Ll}+T_{1R}Z_R^*g_{Rl})+T_{1L}Z_L^*g_{Rl}+T_{1R}Z_R^*g_{Ll}]\nonumber\\
&+\frac23\,\mathrm{Re}[-4\, (T_{2R}Z_L^*g_{Ll}+T_{2L}Z_R^*g_{Rl})-2(T_{2L}Z_R^*g_{Ll}+T_{2R}Z_L^*g_{Rl})]\bigg\}. 
\end{align}
A slight modification of the above expression is required  when there are two different lepton flavors in the final state \cite{Abada:2014kba, Mandal:2019gff}:
\begin{align}
Br(\ell_l^- \to \ell_m^- \ell_n^- \ell_n^+)=
&\frac{\alpha_e^2m_{\ell_i}^5}{32\pi\Gamma_{\ell_i}}\;
\bigg\{ \frac23 (|T_{1L}|^2+|T_{1R}|^2)+\left(|T_{2L}|^2+|T_{2R}|^2\right)\left(\frac{16}{3}\ln\frac{m_{\ell_i}}{m_{\ell_n}}-8\right)\nonumber\\
&-\frac83\,\mathrm{Re}[T_{1L}T_{2R}^*+T_{2L}T_{1R}^*]
+\frac13\left(|Z_L g_{Ll}|^2+|Z_R g_{Rl}|^2+|Z_L g_{Rl}|^2+|Z_R g_{Ll}|^2\right)\nonumber\\
&+\frac23\,\mathrm{Re}[ T_{1L}Z_L^*g_{Ll}+T_{1R}Z_R^*g_{Rl}+T_{1L}Z_L^*g_{Rl}+T_{1R}Z_R^*g_{Ll}]\nonumber\\
&-\frac43\,\mathrm{Re}[T_{2R}Z_L^*g_{Ll}+T_{2L}Z_R^*g_{Rl}+T_{2L}Z_R^*g_{Ll}+T_{2R}Z_L^*g_{Rl}]\bigg\}. 
\end{align}
Photon ($Z$-boson) penguin-diagrams are encoded in the $T_{1L,1R}$ and $T_{2L,2R}$ ($Z_{L,R}$) terms. We derive the following expressions of these terms from $S_1$ and $S_3$ LQs:
\begin{align}
&T^{S_1,S_3}_{1L}=0,
\\
&T^{S_1}_{1R}=\frac{-3}{16\pi^2M^2_1} (Vy^L)_{ql}(Vy^L)^*_{qm}  \left[ \left( \frac{4}{9}+\frac{1}{3}\log(x_q) \right)\frac{2}{3}-\frac{1}{54} \right],
\\
&T^{S_3}_{1R}=\frac{-3}{16\pi^2M^2_3} \bigg\{(Vy^S)_{ql}(Vy^S)^*_{qm}  \left[ \left( \frac{4}{9}+\frac{1}{3}\log(x_q) \right)\frac{2}{3}-\frac{1}{54} \right]
\bigg\},
\\
&T^{S_1}_{2L}=\frac{-3}{16\pi^2M^2_1} \bigg\{
\left[\frac{1}{6}(Vy^L)_{ql}(Vy^L)^*_{qm} +\frac{m_q}{m_l} (Vy^L)_{ql}(y^R)^*_{qm}
\left( \frac{3}{2}+\frac{1}{3}\log(x_q) \right) \right] \frac{2}{3}
\nonumber \\& \hspace{3cm}+
\left[\frac{1}{12}(Vy^L)_{ql}(Vy^L)^*_{qm} -\frac{1}{2}\frac{m_q}{m_l} (Vy^L)_{ql}(y^R)^*_{qm}\right] \frac{-1}{3}
\bigg\},
\\
&T^{S_1}_{2R}=\frac{-3}{16\pi^2M^2_1} \bigg\{
\left[\frac{m_q}{m_l} y^R_{ql}(Vy^L)^*_{qm}
\left( \frac{3}{2}+\frac{1}{3}\log(x_q) \right) \right] \frac{2}{3}
+
\left[ -\frac{1}{2}\frac{m_q}{m_l} y^R_{ql}(Vy^L)^*_{qm}\right] \frac{-1}{3}
\bigg\},
\\
&T^{S_3}_{2L}= \frac{-3}{16\pi^2M^2_3} \bigg\{\frac{1}{12} (Vy^S)_{ql}(Vy^S)^*_{qm}
-\frac{1}{3}y^S_{ql}(y^S)^*_{qm} \bigg\}
\\
&T^{S_3}_{2R}=0.
\end{align}
\begin{align}
&Z^{S_1,S_3}_L=0,
\\    
&Z^{S_1}_R=\frac{-3}{16\pi^2M^2_1}\frac{(Vy^L)_{ql}(Vy^L)^*_{qm}}{m^2_Z \sin^2\theta_W \cos^2\theta_W} \bigg\{
\frac{3}{4}m^2_lg_{u_R}-m^2_q\left(1+\log(x_q)\right)g_{u_L}-\frac{3}{4}m^2_lg_{S_1}
\bigg\},
\\
&Z^{S_3}_R=\frac{-3}{16\pi^2M^2_3}\frac{1}{m^2_Z \sin^2\theta_W \cos^2\theta_W} \bigg\{
\left[ \frac{3}{4}m^2_lg_{d_R}-m^2_q\left(1+\log(x_q)\right)g_{d_L}-\frac{3}{4}m^2_lg^d_{S_3} \right] y^S_{ql}(y^S)^*_{qm}
\nonumber \\ & \hspace{3cm}+
\left[ \frac{3}{4}m^2_lg_{u_R}-m^2_q\left(1+\log(x_q)\right)g_{u_L}-\frac{3}{4}m^2_lg^u_{S_3} \right] (Vy^S)_{ql}(Vy^S)^*_{qm}
\bigg\}.
\end{align}
Here we have defined: $g_{f_L}= I^f_3-Q^f\sin^2\theta_W$, $g_{f_R}=-Q^f\sin^2\theta_W$, $g_{S_1}=\sin^2\theta_W/3= g^u_{S_3}$, and $g^d_{S_3}=4\sin^2\theta_W/3$. Moreover, $\theta_W$ is the Weinberg angle. 

The current experimental bounds of these processes are quoted below  \cite{Hayasaka:2010np, Bellgardt:1987du}:
\begin{align}
&Br(\mu^{\pm}\to e^{\pm}e^+e^-) < 1.0\times 10^{-12},\\
&Br(\tau^{\pm}\to \mu^{\pm}\mu^+\mu^-) < 2.1\times 10^{-8},\\
&Br(\tau^{\pm}\to \mu^{\pm}e^+e^-) < 1.5\times 10^{-8}.
\end{align}

\subsection{\texorpdfstring{$Z\to \ell \ell^{\prime}$}{Z to ll} Processes}
The $Z$-boson decays to leptons receive contributions from the LQs that constraint the Yukawa couplings. These processes are explained with the following effective Lagrangian:
\begin{align}
\delta \mathcal{L}^{Z\to \ell \ell^{\prime}}_{eff}=\frac{g}{\cos \theta_W} \sum_{f,i,j} \overline{f}\gamma^{\mu} \left( g^{ij}_{f_L}P_L + g^{ij}_{f_R}P_R \right) f_j Z_{\mu}.   
\end{align}
Here $g$ is the $SU(2)_L$ gauge coupling. These dimensionless couplings $g^{ij}$ are very accurately measured  at the LEP   \cite{Tanabashi:2018oca} that provide stringent constraints on the associated Yukawa couplings for a fixed LQ mass. NP contributions to these dimensionless couplings can be expressed as follows \cite{Arnan:2019olv}:
\begin{align}
Re\left[ \delta g^\ell_{L,R} \right]^{ij} &=  \dfrac{3w^u_{tj}(w^u_{ti})^{\ast}}{16 \pi^2} \Bigg{[} (g_{u_{L,R}}-g_{u_{R,L}})\dfrac{x_t (x_t-1- \log x_t)}{(x_t-1)^2} \Bigg{]}\nonumber \\
&+ \dfrac{x_Z}{16\pi^2} \sum_{q=u,c}w^u_{qj}(w^u_{qi})^{\ast}  \Bigg{[} g_{u_{L,R}}\left( \log x_Z-\frac{1}{6} \right)+\frac{g_{\ell_{L,R}}}{6} \Bigg{]}\nonumber \\
&+\dfrac{x_Z}{16\pi^2} \sum_{q=d,s,b}w^d_{qj}(w^d_{qi})^{\ast}  \Bigg{[} g_{d_{L,R}}\left( \log x_Z-\frac{1}{6} \right)+\frac{g_{\ell_{L,R}}}{6} \Bigg{]},
\end{align}
When calculating $\delta g_L$  we have defined $w^u_{ij}=(Vy^L)_{ij}$, $w^d_{ij}=0$ ($w^u_{ij}=-(V^*y^S)_{ij}$, $w^d_{ij}=-\sqrt{2}y^S_{ij}$) for $S_1$ ($S_3$) LQ. Similarly while calculating $\delta g_R$ we make the replacements $w^u_{ij}=y^R_{ij}$, $w^d_{ij}=0$ ($w^u_{ij}=0$, $w^d_{ij}=0$) for $S_1$ ($S_3$) LQ. The results from the LEP collaboration \cite{ALEPH:2005ab} provide the following limits on the NP contributions:  
\begin{align}
&Re[\delta g^{ee}_R]\leq 2.9\times 10^{-4},\;\;  Re[\delta g^{\mu\mu}_R]\leq 1.3\times 10^{-3},\;\; Re[\delta g^{\tau\tau}_R]\leq 6.2\times 10^{-4},\\
&Re[\delta g^{ee}_L]\leq 3.0\times 10^{-4},\;\;  Re[\delta g^{\mu\mu}_L]\leq 1.1\times 10^{-3}, \;\;Re[\delta g^{\tau\tau}_L]\leq 5.8\times 10^{-4}.
\end{align}

Furthermore, the  branching ratio for the processes $Z\to \ell \ell^{\prime}$ are given by \cite{Arnan:2019olv}:
\begin{align}
\mathcal{B}(Z\to f_i \bar{f_j}) = \dfrac{m_Z \lambda^{1/2}_Z}{6\pi v^2 \Gamma_Z} \Bigg{[}&\left(|g_{f_L}^{ij}|^2+|g_{f_R}^{ij}|^2\right)\Bigg{(}1- \dfrac{m_{i}^2+m_{j}^2}{2m_Z^2}- \dfrac{(m_i^2-m_j^2)^2}{2m_Z^4}\Bigg{)} \nonumber \\
&+ 6 \dfrac{m_i m_j}{m_Z^2} \mathrm{Re}\left[g_{f_L}^{ij} \left( g_{f_R}^{ij}\right)^\ast\right] \Bigg{]},
\end{align}
here $\lambda_Z\equiv[m_Z^2-(m_i-m_j)^2][m_Z^2-(m_i+m_j)^2]$. Both LEP and LHC results put limits on these branching ratios which are \cite{Akers:1995gz, Abreu:1996mj, Aad:2014bca} summarized below: 
\begin{align}
&Br(Z\to e^{\pm}\mu^{\mp}) < 7.5\times 10^{-7},\\  
&Br(Z\to e^{\pm}\tau^{\mp}) < 9.8\times 10^{-6},\\  
&Br(Z\to \mu^{\pm}\tau^{\mp}) < 1.2\times 10^{-5}.
\end{align}

As for the neutrinos, $Z$-decays of the form $Z\to \nu \nu$ 
also receive contributions form LQs that are parametrized by,
\begin{align}
N_\nu = \sum_{i,j} \left|\delta_{ij}+\dfrac{\delta g_{\nu_L}^{ij}}{g_{\nu_L}^{\mathrm{SM}}}\right|^2,
\;\;\;N_\nu^{\mathrm{exp}} = 2.9840\pm 0.0082 \text{\cite{ALEPH:2005ab}}.
\end{align}
Above we have also collected the accurately measured experimental value of this effective number of neutrinos.

\subsection{\texorpdfstring{$\mu - e$}{mu-e} Conversion} 
With the choice of the Yukawa coupling matrices given in Eq. \eqref{matrix}, $S_3$ LQ mediates $\mu - e$ conversion in nuclei at the tree-level in our model. This conversion rate can be calculated from the following formula \cite{Kitano:2002mt,Dorsner:2016wpm, Saad:2020ucl}:     
\begin{align}
&CR(\mu-e) 
= \frac{\Gamma^{\mu-e} }{\Gamma_\text{capture}(Z)},
\\
&\Gamma^{\mu-e} \,= \,2 \,G_F^2\, 
\left| (2 V^{(p)} + g_{LV}^{(u)} V^{(n)}) g_{LV}^{(u)}\right|^2,
\\
&g_{LV}^{(u)}=\frac{-2v^2}{m_{S_3}^2} \,  
(V^{\ast} y^S)_{u\ell'}\, (V^{\ast}y^S)^*_{u\ell}.
\end{align}
$\Gamma_\text{capture}(Z)$ is the total capture rate for a nucleus with atomic number $Z$, which is $13.07\times 10^6$ $s^{-1}$ for gold, and the corresponding nuclear form factors  are given by \cite{Kitano:2002mt}  $V^{(p)}=0.0974$,  $V^{(n)}=0.146$ (in units of $m_{\mu}^{5/2}$). The current  sensitivity implies $\text{CR}(\mu-e)<7\times 10^{-13}$   \cite{Bertl:2006up}, whereas  the future projected sensitivity is expected to make almost four orders
of magnitude improvement over the current limit  $CR(\mu\to e)< 10^{-16}$  \cite{Kurup:2011zza, Cui:2009zz, Chang:2000ac, Adamov:2018vin, Bartoszek:2014mya, Pezzullo:2018fzp, Bonventre:2019grv}.

\subsection{\texorpdfstring{$P^0\to\ell^-\ell^{\prime+}$}{meson to ll}} 
For the explanations of the $R_{K^{(*)}}$ ratios along with neutrino oscillation data, the NP contributions to the $O_{9,10}$ operators need to be large. The associated Wilson coefficients  $C_{9,10}$ as given in Eq. \eqref{RK1} then lead to interesting pseudoscalar meson decays via $b\to s \mu^+\mu^-$, $b\to s \mu^+\tau^-$, and $b\to s \tau^+\tau^-$. The decay width of the process $P^0 \to \ell^- \ell^{\prime+}$ can be written as \cite{Becirevic:2016zri}:
\begin{align}
\Gamma_{P \to \ell^{-}\ell^{\prime+}} &= 
f_{P}^2 m_{P}^3 \frac{G_F^2 \,\alpha_{e}^2}{64\pi^3}  
\left|V_{qj}V_{qi}^*\right|^2
\eta(m_{P},m_\ell,m_{\ell^{\prime}})
\scalemath{0.85}{
\left(
\left|
\frac{(m_\ell-m_{\ell^{\prime}})}{m_{P}} 
\left(C^{ij;\ell \ell^{\prime}}_{9}\right)
\right|^2 +\left| 
\frac{(m_\ell+m_{\ell^{\prime}})}{m_{P}} 
\left(C^{ij;\ell \ell^{\prime}}_{10}\right)
\right|^2 
\right)
}.
\end{align}
For our scenario $B_s$ is the only relevant meson, which  corresponds to $q=t, j=b, i=s$ in the above formula, and the function $\eta$ is defined as: 
\begin{align}
\eta(m_{P},m_\ell,m_{\ell^{\prime}}) = 
\sqrt{[1-(m_\ell-m_{\ell^{\prime}})^2/m_{P}^2] 
[1-(m_\ell+m_{\ell^{\prime}})^2/m_{P}^2)]}.
\end{align}
The experimental limits on these processes are given below  \cite{Aaij:2019okb, Aaij:2017xqt, Aaij:2017vad}:
\begin{align}
&Br(B_s\to \mu^{\pm} \mu^{\mp})_{\text{exp}}= (3.0\pm 0.6)\times 10^{-9},
\\
&Br(B_s\to \mu^{\pm} \tau^{\mp})_{\text{exp}} \leq 4.2\times 10^{-5},
\\
&Br(B_s\to \tau^{\pm} \tau^{\mp})_{\text{exp}}\leq 6.8\times 10^{-3}.
\end{align}
Among these, only  $B_s \to \mu^+ \mu^-$ decay mode has been observed, which is in good agreement with the   SM prediction \cite{Bobeth:2013uxa}, $Br(B_s \to \mu^+ \mu^-)_{\text{SM}}=(3.65\pm 0.23)\times 10^{-9}$.

Associated to $b\to s \mu^+\tau^-$ transition there is another important constraint that comes from $B\to K$ decay that has the following branching ratio \cite{Crivellin:2015era}:
\begin{align}
Br(B^+\to K^+ \tau^{\pm} \mu^{\mp})= \left\{ 9.6 \left( |C^{23}_9|^2 +|C^{32}_9|^2 \right) + 10 \left( |C^{23}_{10}|^2 +|C^{32}_{10}|^2 \right) \right\}\times 10^{-9},    
\end{align}
with the following experimental bound on this process \cite{Lees:2012zz}:
\begin{align}
    Br(B^+\to K^+ \tau^{\pm} \mu^{\mp}) \leq 4.8\times 10^{-5}.
\end{align}

\subsection{\texorpdfstring{$B\to K^{(*)}\nu \overline{\nu}$}{B to K nu nu}}
Both $S_1$ and $S_3$ LQs can induce $B\to K^{(*)}\nu \overline{\nu}$ decay at the tree-level via $d_k\to d_j\nu\overline{\nu}$ processes.  The Wilson coefficients responsible for such decays associated to $b\to s$ transition are:
\begin{align}
C^{fi}_L= \frac{\pi v^2}{2 V_{tb}V^*_{ts}\alpha} \left\{\frac{y^L_{bi}(y^L)^*_{sf}}{M^2_1} + \frac{y^S_{bi}(y^S)^*_{sf}}{M^2_3} \right\}.  
\end{align}
Then following  \cite{Buras:2014fpa}, the branching ratio for $B\to K^{(*)}\nu \overline{\nu}$ can be expressed as:
\begin{align}
R^{\nu \overline{\nu}}_{K^{(*)}}= \frac{1}{3 |C^{\text{SM}}_L|^2} \sum_{i,f=1}^3 \left|  \delta^{fi}C^{\text{SM}}_L +C^{fi}_L \right|^2,   
\end{align}
this ratio is normalized to SM, where  $C^{\text{SM}}_L=-1.47/\sin^2\theta_W$. The Belle collaboration \cite{Grygier:2017tzo} limits these ratios to be  $R^{\nu \overline{\nu}}_K < 3.9$ and $R^{\nu \overline{\nu}}_{K^*} < 2.7$. 

\subsection{\texorpdfstring{$B_c\to \tau \nu$}{B to tau nu}} 
The same Wilson coefficients that explain $R_{D^{(*)}}$ in our framework also lead to   $B_c\to \tau \nu$ decay. The associated branching ratio that depends on the vector and the scalar Wilson coefficients can be written as \cite{Watanabe:2017mip, Blanke:2018yud}: 
\begin{align}
Br(B_c\to \tau \nu)=0.023 \left|1+C^{33}_V-4.3\; C^{33}_S\right|^2.
\end{align}

The lifetime of $B_c$ has not been measured in the experiments yet. Hence, this quantity needs to be compared with the theoretical calculations \cite{Gershtein:1994jw, Bigi:1995fs,Beneke:1996xe, Chang:2000ac,  Kiselev:2000pp}. By carrying out such calculations in Ref. \cite{Akeroyd:2017mhr} and Ref. \cite{Alonso:2016oyd}, their results advocate that the NP contributions to this decay must be $Br(B_c\to \tau \nu)\leq  10\%$ and $Br(B_c\to \tau \nu)\leq 30\%$, respectively. On the other hand, as argued in  Refs. \cite{Blanke:2018yud, Bardhan:2019ljo}, these calculations suffer from theoretical uncertainties, and suggested a conservative limit of $Br(B_c\to \tau \nu)\leq 60\%$. It is interesting to note that $R_2$ LQ explanations to $R_{D^{(*)}}$ demands much larger values \cite{Saad:2020ucl} of this branching ratio, hence such explanations can in principle  be ruled out by reducing the corresponding uncertainties in future. On the contrary, the observation of $R_{D^{(*)}}$ can be properly accommodated via $S_1$ LQ  with  smaller values of this branching ratio \cite{Crivellin:2019dwb}.

\subsection{\texorpdfstring{$\tau \to \ell P^0$}{tau decay} Decays}
In the SM tau lepton decays into mesons and lighter leptons are not allowed. However, these lepton flavor violating decays can be significant in the presence of leptoquarks that provide strong constraints on the Yukawa couplings.  Tau lepton decay width for $\tau \to \ell P^0$ process can be written as follows \cite{Mandal:2019gff}: 
\begin{align}
\Gamma (\tau \to \ell P^0)= \frac{f^2_P\lambda^{1/2}_P}{128\pi m^3_{\tau}} \left[ \left( m^2_{\tau} +m^2_{\ell} -m^2_P \right) \left( |\alpha_P|^2 +|\beta_P|^2 \right) + m_{\tau} m_{\ell} Re (\alpha_P \beta_P) \right],  \label{tau}  
\end{align}
where $\lambda (a,b,c)=a^2+b^2+c^2-2ab-2ac-2bc$. Within our scenario, the related processes we need to take into account  are for $P=\phi, \eta, \eta'$. As for the meson form factors we take the number quoted in Ref. \cite{Bhattacharya:2016mcc}, and their masses are taken from Ref. \cite{Tanabashi:2018oca}. From hereafter, we will neglect the mass of the lighter charged lepton. With this assumption, the only relevant terms that enter in Eq. \eqref{tau} are:
\begin{align}
\alpha_{\phi}=& \frac{m_\tau}{M^2_3} y^S_{23}(y^S_{22})^*,
\\
\alpha_{\eta}=& \frac{m_\tau}{2\sqrt{3}} \left(
-\frac{1}{M^2_1}(V^*y^L)_{13}(V^*y^L)^*_{12}
-\frac{1}{M^2_3}(V^*y^S)_{13}(V^*y^S)^*_{12} \right.
\nonumber \\& \left.
\hspace{100pt} +\frac{1}{M^2_1}(V^*y^L)_{23}(V^*y^L)^*_{22}
+\frac{2}{M^2_3}y^S_{23}(y^S_{22})^*
\right),
\\
\alpha_{\eta'}=& \frac{m_\tau}{2\sqrt{6}} \left(
\frac{1}{M^2_1}(V^*y^L)_{13}(V^*y^L)^*_{12}
+\frac{1}{M^2_3}(V^*y^S)_{13}(V^*y^S)^*_{12} \right.
\nonumber \\& \left.
\hspace{100pt} +\frac{2}{M^2_1}(V^*y^L)_{23}(V^*y^L)^*_{22}
+\frac{4}{M^2_3}y^S_{23}(y^S_{22})^*
\right).
\end{align}

Current bounds on these processes are  \cite{Tanabashi:2018oca}, 
\begin{align}
&Br(\tau \to \mu \phi) \leq 8.4\times 10^{-8},\\ 
&Br(\tau \to \mu \eta) \leq 6.5\times 10^{-8},\\
&Br(\tau \to \mu \eta') \leq 1.3\times 10^{-7}.\\
\end{align}

\subsection{\texorpdfstring{$B^0_s - \overline{B^0_s}$}{meson mixing} Mixing} 
Concerning the LQs, both $S_1$ and $S_3$ contribute to meson-antimeson mixing. This NP contribution to $B^0_s - \overline{B^0_s}$ mixing can be described by the effective Lagrangian given below \cite{Marzocca:2018wcf}:
\begin{align}
\mathcal{L}^{\Delta B=2}_{eff}= -(C^{\text{SM}}_1+C^{NP}_1)  \left( \overline{b}_L \gamma_\mu s_L  \right)^2.   
\end{align}
Here the SM part is $C^{\text{SM}}_1= 2.35/(4\pi^2)\; \left( V_{tb}V^*_{ts} G_F m_W \right)^2$ \cite{Lenz:2010gu} and the NP contribution at the heavy scale ($\Lambda$) is given by \cite{Dorsner:2016wpm, Bobeth:2017ecx, Marzocca:2018wcf, Crivellin:2019dwb}:
\begin{align}
C^{NP}_1= \frac{1}{128\pi^2} \left\{ \frac{1}{M^2_1} \left[(y^L)^*_{2i}y^L_{3i}\right]^2+  \frac{5}{M^2_3} \left[(y^S)^*_{2i}y^S_{3i}\right]^2+ \frac{2}{M_1M_3}\left[(y^L)^*_{2i}y^L_{3i}\right] \left[(y^S)^*_{2j}y^S_{3j}\right] \right\}.
\end{align}
Here we neglect the evolution of $C^{NP}_1$ from high scale to the $m_w$ scale. Then the mass difference is given by:
\begin{align}
\Delta m_{B_s}^{SM+NP} = \Delta m_{B_s}^{\text{SM}}\left| 1+\frac{C_1^{NP}}{C_1^{\text{SM}}}  \right|.    
\end{align}
The SM prediction is $\Delta m_{B_s}^{\text{SM}}= (18.3 \pm 2.7)\times 10^{12}s^{-1}$ \cite{Bona:2006sa, Jubb:2016mvq}. This mass difference has been measured in the experiments \cite{Bona:2008jn, Tanabashi:2018oca} with great accuracy, which is given by:
\begin{align}
\Delta m_{B_s}^{\text{exp}}=  (17.757 \pm 0.021)\times 10^{12}s^{-1}. 
\end{align}

\section{Numerical Analysis and Discussion}\label{SEC-05}
In this section we perform a numerical analysis of the proposed model to demonstrate how to reconcile neutrino oscillation data with anomalies in the B meson decays and  the muon anomalous  magnetic moment. 

\begin{figure}[b!]
\centering
\includegraphics[scale=0.4]{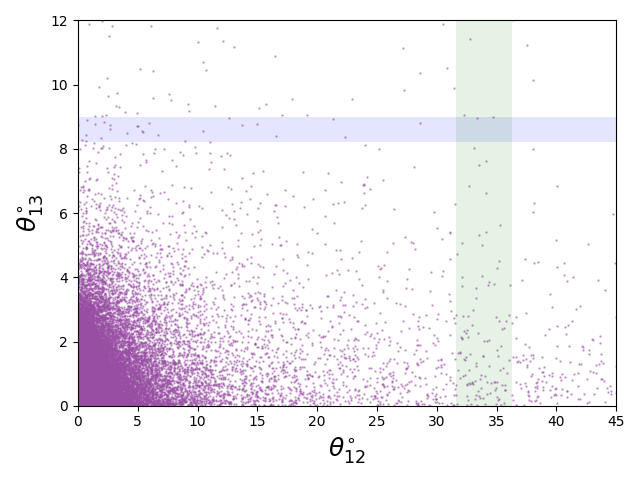}\hspace{0.8cm}
\includegraphics[scale=0.4]{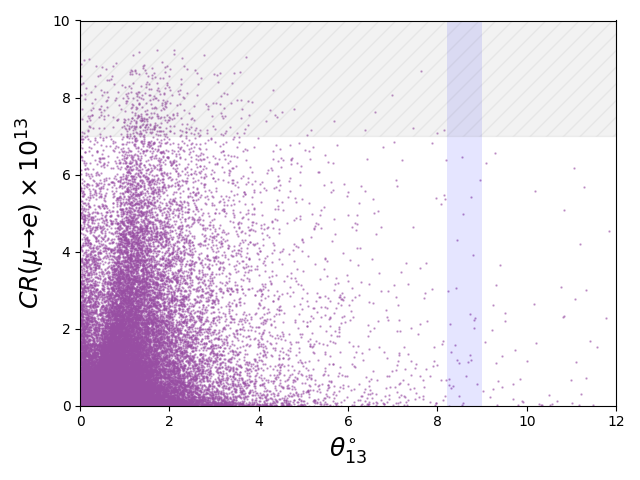}
\caption{ The results of random scans showing the correlations  between $\theta_{13}$ and $\theta_{12}$ on the left plot, $CR(\mu\to e)$ and $\theta_{13}$ on the right plot, respectively. In making these plots, we have randomly varied the relevant couplings fifteen thousand times in  ranges between:  $y^L_{22,33}, y^S_{33}, y^{\omega}_{23}=[-0.1,0.1]$,  $y^L_{23}, y^{\omega}_{22}=[-1.5,1.5]$, $y^L_{32}, y^{\omega}_{33}=[-0.05,0.05]$, $y^S_{22,23,32}=[-0.5,0.5]$, and $y^S_{31}=[-0.01,0.01]$. The shaded blue (green) region corresponds to $3\sigma$ allowed  values of $\theta_{13}$ ($\theta_{12}$). Moreover, in all plots hatched gray area represents experimental exclusion region of the corresponding  quantity.} \label{plot1}
\end{figure}
First we recall that for simplicity, we consider all parameters of this theory to be real. Extension to the general case with complex Yukawa couplings is straightforward. In this CP-conserving scenario, we adopt the Wolfenstein parametrization \cite{Wolfenstein:1983yz} for the CKM matrix:
\begin{align}
V=\begin{pmatrix}
1-\frac{1}{2}\lambda^2&\lambda&A\lambda^3\rho\\
-\lambda&1-\frac{1}{2}\lambda^2&A\lambda^2\\
A\lambda^3(1-\rho)&-A\lambda^2&1
\end{pmatrix},
\end{align}
and take values of the mixing parameters $\lambda$=0.2248, A=0.8235, $\rho$=0.1569  \cite{Tanabashi:2018oca}. Masses of the down-quarks enter into the computation of the neutrino mass matrix and we take their values to be  $m_d=4.7$ MeV, $m_s=95$ MeV and $m_d=4.18$ GeV  \cite{Tanabashi:2018oca}. Furthermore, for this numerical study done in this section, we choose $M_1=M_3=0.12 M_{\omega}$ and set $M_{\omega}=10$ TeV. However, masses of these LQs need not be degenerate in general. The masses of the leptoquarks and diquark chosen here are consistent with collider bounds, which we will discuss shortly.  With degenerate LQ masses and by further assuming $\mu_1=\mu_3=\mu$ just for simplicity, the neutrino mass formula given in Eq. \eqref{numatrix}  can be written as
\begin{align}
\mathcal{M}^{\nu}_{ij} = m_0 \;y^p_{li}\; m^d_{ll}\; y^{\omega}_{lk}\;  m^d_{kk}\; y^p_{kj},\;\;\; m_0= 3\mu/(32\pi^4 M^2_{LQ}) \mathcal{\overline{I}}\left[M^2_{DQ}/M^2_{LQ}\right].
\end{align}
This corresponds to $m_0=1.95\times 10^{-10} \mu$, then $\mu$ can be fixed from one of the two measured neutrino mass squared differences.

\begin{figure}[t!]
\centering
\includegraphics[scale=0.4]{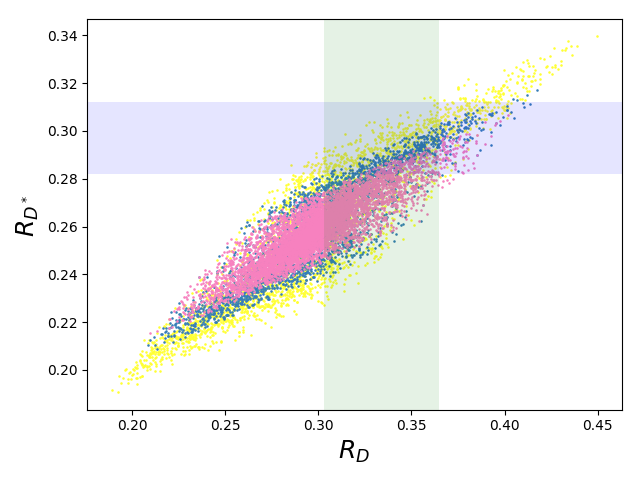}\hspace{0.8cm}
\includegraphics[scale=0.4]{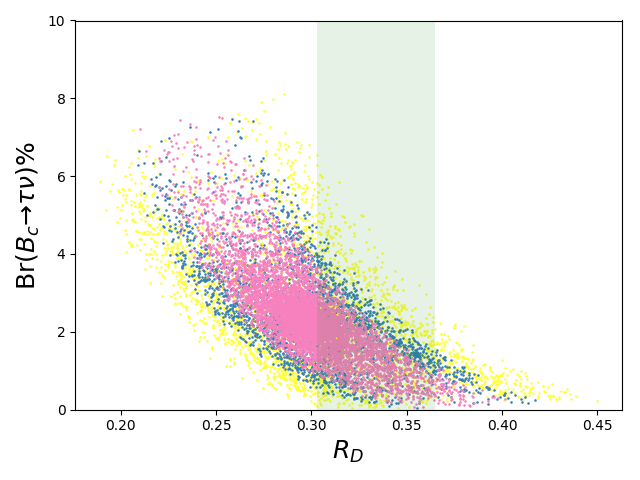}
\caption{ The results of random scans showing the correlations  between $R_{D^*}$ and $R_D$ on the left plot, $Br(B_c\to \tau \nu)$ and $R_D$ on the  right plot, respectively. Pink dots correspond to scenario where new physics contribution to $B^0_s-\overline{B^0_s}$ mixing $< 10\%$. Similarly, blue (yellow) dots correspond to scenario where new physics contribution to $B^0_s-\overline{B^0_s}$ mixing is in between $10\%$ and $20\%$ ($20\%$ and $50\%$).  In making these plots, we have randomly varied the relevant couplings in  ranges between: $y^{L,R}_{23}=[-1.7,1.7]$, $y^{L,S}_{23}=[-0.5,0.5]$, and  $y^{L,S}_{33}=[-0.25,0.25]$. The horizontal (vertical) shaded blue (green) region corresponds to $1\sigma$ values of $R_{D^*}$ ($R_D$).} \label{plot2}
\end{figure}
\begin{figure}[t!]
\centering
\includegraphics[scale=0.4]{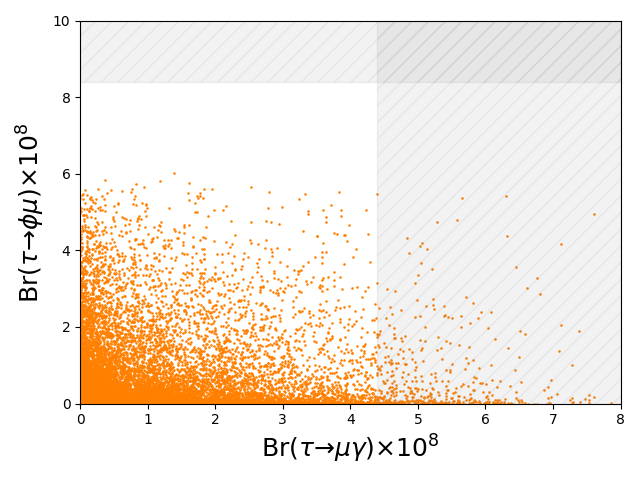}\hspace{0.8cm}
\includegraphics[scale=0.4]{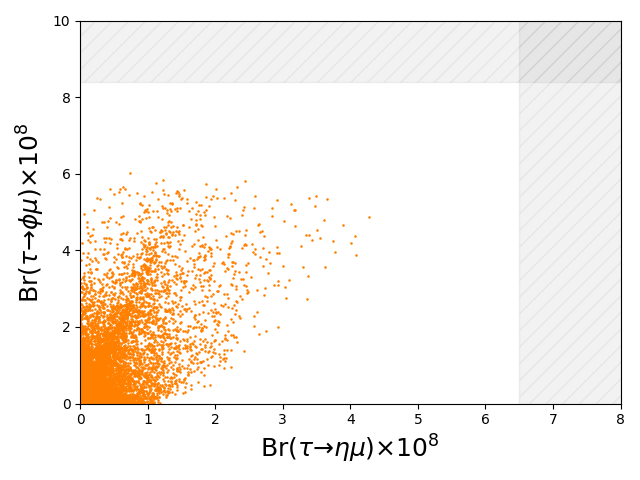}
\caption{ The results of random scans showing the interrelations  between $Br(\tau\to \phi\mu)$ and $Br(\tau\to \mu\gamma)$ on the  left plot, and $Br(\tau\to \phi\mu)$ and $Br(\tau\to \mu\gamma)$ on the right plot, respectively.  In making these plots, we have randomly varied the relevant couplings in  ranges between: $y^{R,L}_{32}, y^S_{22}=[-0.05,0.05]$, $y^L_{22}, y^S_{33}=[-0.1,0.1]$,  $y^{L,S}_{23}=[-0.5,0.5]$, $y^S_{32}=[-0.2,0.2]$, and $y^L_{33}=[-0.006,0.006]$.} \label{plot3}
\end{figure}
\begin{figure}[t!]
\centering
\includegraphics[scale=0.4]{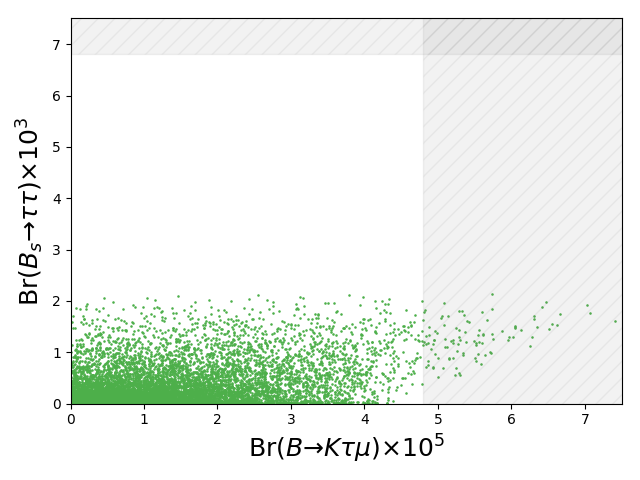}\hspace{0.8cm}
\includegraphics[scale=0.4]{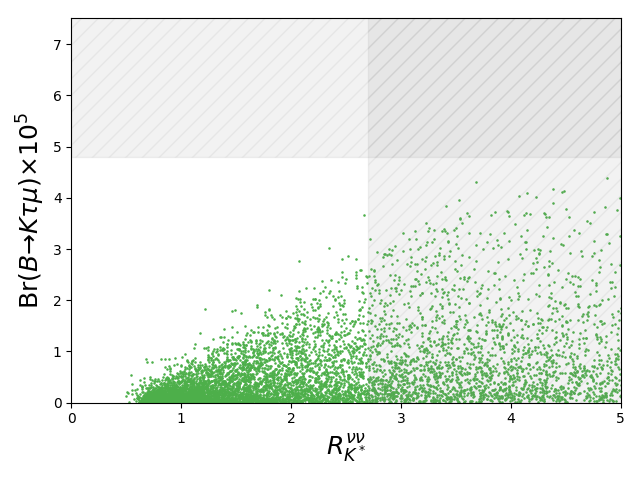}
\caption{ The results of random scans showing the links  between $Br(B_s\to \tau \tau)$ and $Br(B_s\to K\tau \mu)$ on the left plot, $Br(B_s\to K\tau \mu)$ and $R^{\nu\nu}_{K^*}$ on the right plot, respectively.  In making these plot, we have randomly varied the relevant couplings in  ranges between: $y^{L,S}_{22}=[-0.1,0.1]$, $y^L_{23}, y^S_{23,32}=[-0.5,0.5]$, $y^L_{32,33}=[-0.05,0.05]$, and $y^S_{33}=[-1,1]$.} \label{plot4}
\end{figure}

As for the neutrinos alone it is trivial to get a fit to the data from the above mass matrix formula. However, as elaborated in the previous sections,  the explanations of the muon $g-2$ puzzle, $R_K$, $R_{K^*}$ and $R_D$, $R_{D^*}$ flavor anomalies as well as neutrino masses and mixings are all directly intertwined with each other.  Moreover, the same set of Yukawa couplings also  leads to many other flavor violating processes as described in Sec. \ref{SEC-04}. This makes our scenario both challenging and attractive at the same time.  For illustrations we present some of these correlations among different physical quantities in Figs. \ref{plot1}, \ref{plot2}, \ref{plot3}, and \ref{plot4}.

The type of  Yukawa coupling textures that we consider in this work is already introduced in Eq. \eqref{matrix}. As we have  discussed in Sec. \ref{SEC-03}, in a scenario with only entries in the lower $2\times 2$ blocks for all the matrices does not lead to realistic neutrino fit. It is trivial to understand that two of the three mixings angles $\theta_{12}$ and $\theta_{13}$ would remain zero in this case. As we have argued, to alleviate this issue, one needs to consider at least one non-vanishing term among $\{y^{S}_{21}, y^{S}_{31}, y^{S}_{21}, y^{S}_{31}\}$, and we have made an ad hoc choice of  $y^{S}_{31}\neq 0$ just for demonstration. An immediate consequence is that non-zero $y^{S}_{31}$ leads to $\mu \to e$ conversion in the nuclei. Hence, neutrino oscillations are directly linked to lepton flavor violating processes. Correlations among these quantities are depicted in Fig. \ref{plot1} by randomly varying the relevant Yukawa couplings.

Since within our scenario, both the vector and scalar-tensor Wilson coefficients take part in explaining the $R_{D^{(*)}}$ ratios, significant new physics contributions to $B^0_s-\overline{B^0_s}$ mixing, as well as in $B\to K^{(*)}\nu\nu$ process are unavoidable. 
This situation is illustrated by showing the interdependence between $R_D$ and $R_{D^*}$ in Fig. \ref{plot2} (left plot). Here, pink dots correspond to a scenario where NP contribution to $B^0_s-\overline{B^0_s}$ mixing $< 10\%$. Similarly, blue (yellow) dots the case where NP contribution to $B^0_s-\overline{B^0_s}$ mixing is in between $10\%$ and $20\%$ ($20\%$ and $50\%$). From this plot, it is clear that a fit to both $R_D$ and $R_{D^*}$ to their experimental central values require more than $10\%$ contribution to $B^0_s-\overline{B^0_s}$ mixing.

In the same figure, the plot on the right shows the interrelationship between $R_D$ and the branching ratio for $B_c\to \tau \nu$. As can be seen from this plot that correct values of $R_D$ and $R_{D^*}$ ratios can be reproduced within this set-up even with  $Br(B_c\to \tau \nu) < 10\%$, which is unlike the scenarios when $R_2$ LQ is employed to explain B meson decay anomalies in the charged current processes that demands large branching ratio of this process (see for example Ref. \cite{Saad:2020ucl}). Another immediate difference between utilizing $R_2$ and $S_1$ that we point out here is, even though in our scenario  for certain choices of parameters, NP contributions to $Z\to \tau_{L(R)}\tau_{L(R)}$ can be large, consistent fits can be obtained where these relevant contributions are small (see Table \ref{result}). However, when $S_1$ is replaced with $R_2$ LQ, NP effects on $Z\to \tau_{L(R)}\tau_{L(R)}$ decays are usually significant that puts strong restrictions on the upper limit on the associated Yukawa couplings (see for example Ref. \cite{Saad:2020ucl}).

The essential parameters that describe the muon AMM and B-physics  anomalies, as well as neutrino oscillation data unavoidably lead to charged lepton and meson decays. In our set-up, the tau decays to a muon and a photon is the most constraining process. In fact, as long as $\tau\to \mu\gamma$ decay limit is satisfied, $\tau\to \mu\mu\mu$ processes are under control. Interconnections among tau decays to lighter leptons and a photon, as well as its decays to meson and lighter leptons are presented in Fig. \ref{plot3} by varying the relevant Yukawa couplings.  Further correlations among different meson decay modes are portrayed in Fig. \ref{plot4}. 

From our detailed numerical analysis  we find that points that satisfy all fit requirements,  branching fractions for $\tau \to \mu \gamma$, $\tau \to \phi\mu$ and $\tau \to \eta\mu$ are always very close to the current experiment upper limits (see Table \ref{result}). As for the lepton flavor violating $B$ meson decays, $B_s\to \tau \mu$ is expected to be within one or two orders below the current experimental limit, whereas for decays of the form $B\to K\tau\mu$, the expected branching ratios are just one order smaller than the current bounds (see Table \ref{result}). Furthermore, NP contributions to the branching ratios of  $B_s\to \tau \tau$ and  $B_s\to \mu \mu$ are about hundred times enhanced and suppressed, respectively  compared to the SM predictions (see Table \ref{result}). Some of these enhanced effects, such as in $B\to K\tau\mu$ and $\tau\to \phi \mu$ can be tested soon by LHCb and Belle-II collaborations.

For illustration purpose, we also provide concrete benchmark points that incorporate neutrino masses and mixings, as well as accommodate anomalies in the $a_\mu, R_{K^{(*)}}, R_{D^{(*)}}$, and simultaneously  satisfy all experimental constraints. Two such benchmark points are given below: 
\begin{align}
\tt{BM-I:}\label{BM-I}&\\
&y^L=\begin{pmatrix}
0&0&0\\
0&-0.09485&-1.413\\
0&0.01699&-0.05935
\end{pmatrix},\;\;\;
y^R= \begin{pmatrix}
0&0&0\\
0&0&1.451\\
0&0.1900&0
\end{pmatrix},\nonumber \\
&y^S=\begin{pmatrix}
0&0&0\\
0&0.03230&-0.4183\\
0.002867&0.03398&0.1742
\end{pmatrix},\;\;\;
y^{\omega}=\begin{pmatrix}
0&0&0\\
0&-1.451&0.1332\\
0&0.1332&0.04726  
\end{pmatrix}.\nonumber
\end{align}
\newpage
\begin{align}
\tt{BM-II:}\label{BM-II}&\\
&y^L=\begin{pmatrix}
0&0&0\\
0&-0.03947&-1.337\\
0&0.01907&-0.05912
\end{pmatrix},\;\;\;
y^R= \begin{pmatrix}
0&0&0\\
0&0&1.579\\
0&0.1901&0
\end{pmatrix},\nonumber \\
&y^S=\begin{pmatrix}
0&0&0\\
0&-0.06141&0.1807\\
-0.0009881&-0.01793&-0.4249
\end{pmatrix},\;\;\;
y^{\omega}=\begin{pmatrix}
0&0&0\\
0&-0.4746&-1.013\\
0&-1.013&0.002936  
\end{pmatrix}.\nonumber
\end{align}
For \texttt{BM-I} (\texttt{BM-II}) we take $\mu=131.89$ ($\mu=489.13$) GeV. Associated with these benchmark points, values of a long list of observables are tabulated in Table \ref{result}.

\begin{table}[t!]
\centering
\footnotesize
\resizebox{0.7\textwidth}{!}{
\begin{tabular}{c|c|c}
\hline
\textbf{Observables} & \texttt{BM-I}& \textbf{{\tt BM-II}}\\ \hline
\rowcolor{red!10}$\Delta m^2_{21} (eV^2)$&$7.348\times 10^{-5}$&$7.383\times 10^{-5}$\\ 
\rowcolor{red!10}$\Delta m^2_{31} (eV^2)$&$2.526\times 10^{-3}$&$2.524\times 10^{-3}$\\ 
\rowcolor{red!10}$\theta_{12}$&$33.698^{\circ}$& $33.813^{\circ}$\\ 
\rowcolor{red!10}$\theta_{23}$&$48.373^{\circ}$& $48.055^{\circ}$\\ 
\rowcolor{red!10}$\theta_{13}$&$8.624^{\circ}$&$8.615^{\circ}$\\

\rowcolor{yellow!10}$\Delta a_{\mu}$&$2.718\times 10^{-9}$&$2.688\times 10^{-9}$\\ 
\rowcolor{yellow!10}$C_9=-C_{10}$&-0.528&-0.530\\
\rowcolor{yellow!10}$R_D$&0.339&0.340\\ 
\rowcolor{yellow!10}$R_{D^*}$&0.286&0.285\\

\rowcolor{cyan!10}$Br(\mu \to e\gamma)$&$8.284\times 10^{-15}$&$2.740\times 10^{-16}$\\
\rowcolor{cyan!10}$CR(\mu \to e)$&$3.243\times 10^{-16}$&$1.377\times 10^{-16}$\\
\rowcolor{cyan!10}$Br(\tau \to e\gamma)$&$3.862\times 10^{-14}$&$2.730\times 10^{-14}$\\ 
\rowcolor{cyan!10}$Br(\tau \to \mu\gamma)$&$3.860\times 10^{-8}$&$2.750\times 10^{-8}$\\
\rowcolor{cyan!10}$Br(\tau \to \mu\mu\mu)$&$3.844\times 10^{-9}$&$1.147\times 10^{-9}$\\
\rowcolor{cyan!10}$Br(\tau \to \mu e e)$&$3.273\times 10^{-9}$&$8.788\times 10^{-10}$\\

\rowcolor{orange!10}$\delta g_R^{\tau\tau}$&$1.467\times 10^{-4}$&$1.732\times 10^{-4}$\\
\rowcolor{orange!10}$\delta g_L^{\tau\tau}$&$3.955\times 10^{-5}$&$6.094\times 10^{-5}$\\
\rowcolor{orange!10}$Br(Z\to \mu\tau)$&$5.363\times 10^{-15}$&$1.732\times 10^{-12}$\\

\rowcolor{green!10}$Br(B_s\to \mu \mu)$&$4.925\times 10^{-11}$&$4.954\times 10^{-11}$\\
\rowcolor{green!10}$Br(B_s\to \mu \tau)$&$3.803\times 10^{-7}$&$1.125\times 10^{-6}$\\
\rowcolor{green!10}$Br(B_s\to \tau \tau)$&$4.604\times 10^{-5}$&$5.110\times 10^{-5}$\\
\rowcolor{green!10}$Br(B\to K \mu \tau)$&$1.063\times 10^{-6}$&$3.145\times 10^{-6}$\\
\rowcolor{green!10}$R^{\nu\overline{\nu}}_{K^*}$&$1.855$&$1.656$\\
\rowcolor{green!10}$Br(B_c\to \tau \nu)\%$&$1.767$&$1.703$\\

\rowcolor{gray!10}$\Delta m^{NP+SM}_{B_s}/\Delta m^{\text{SM}}_{B_s}$&$1.12$&$1.13$\\

\rowcolor{blue!10}$Br(\tau\to \phi\mu)$&$2.047\times 10^{-8}$&$1.380\times 10^{-8}$\\
\rowcolor{blue!10}$Br(\tau\to \eta\mu)$&$4.397\times 10^{-8}$&$3.123\times 10^{-9}$\\

\end{tabular}
}
\caption{Values of observables associated with the benchmark points given in Eqs. \eqref{BM-I}, \eqref{BM-II}. }\label{result}
\end{table}

\subsection{LHC Bounds}
In this sub-section, we briefly discuss the collider bounds. There exists dedicated direct search for LQs at the LHC that provide strong bounds on the masses of the LQs. From our numerical  inspection the typical types of solutions that we get are of similar forms as the benchmark points presented in Eqs. \eqref{BM-I} and \eqref{BM-II}. This is why it is sufficient to discuss the representative bounds associated with these benchmark points.  Since the Yukawa couplings are not too large in our scenario, hence the main LHC bounds are coming from the QCD driven LQ pair-production. Neglecting the $t$-channel contributions, this corresponds to two different production mechanism: gluon-gluon fusion ($gg\to LQ\overline{LQ}$), and quark-antiquark annihilation ($q\overline{q}\to LQ\overline{LQ}$). Once produced, each LQ will decay into a quark and a lepton, and the bounds on these LQ masses highly depend on the branching fractions to different decay modes.

For illustration let us take for example \texttt{BM-II} of Eq. \eqref{BM-II} to derive these bounds. The decay modes of the LQs are then given by:
\begin{align}
&S^{1/3}_1\to s_L\nu_L (1.3),\; u_L\tau_L (0.3),\; c_L\tau_L (1.3),\; c_R\tau_R (1.5),\; t_R\mu_R (0.2), \label{br1}\\
&S^{4/3}_3\to s_L\tau_L (0.25),\; b_L\tau_L (0.6),\label{br2}\\
&S^{-2/3}_3\to c_L\nu_L (0.22),\; t_L\nu_L (0.61), \label{br3}\\
&S^{1/3}_3\to s_L\nu_L (0.2),\; b_L\nu_L (0.41),\; c_L\tau_L (0.15),\; t_L\tau_L (0.43), \label{br4}
\end{align}
here numbers inside the parentheses are the associated Yukawa couplings responsible for these decay modes. The bounds on the LQ masses for these decay modes from LHC searcher are given as follows: 
\begin{align}
&LQ\overline{LQ}\to jj\nu\overline{\nu}:\; 980\;\text{GeV} \;\;(635\;\text{GeV});
\;\;\; @35.9\;fb^{-1} \;\;\;\text{\cite{Sirunyan:2018kzh}},\\ 
&LQ\overline{LQ}\to b\overline{b}\tau\overline{\tau}:\; 1025\;\text{GeV} \;\;(835\;\text{GeV});
\;\;\; @36.1\;fb^{-1} \;\;\;\text{\cite{Aaboud:2019bye, Sirunyan:2018vhk}},\\ 
&LQ\overline{LQ}\to t\overline{t}\nu\overline{\nu}:\; 1020\;\text{GeV} \;\;(812\;\text{GeV});
\;\;\; @35.9\;fb^{-1} \;\;\;\text{\cite{Sirunyan:2018kzh}},\\ 
&LQ\overline{LQ}\to t\overline{t}\mu\overline{\mu}:\; 1420\;\text{GeV} \;\;(950\;\text{GeV});
\;\;\; @36.1\;fb^{-1} \;\;\;\text{\cite{Camargo-Molina:2018cwu, CMS-PAS-B2G-16-027}},\\ 
&LQ\overline{LQ}\to t\overline{t}\tau\overline{\tau}:\; 930\;\text{GeV} \;\;(730\;\text{GeV});
\;\;\; @36.1\;fb^{-1} \;\;\;\text{\cite{Aaboud:2019bye}}.
\end{align}
Here the current limits on LQ masses are shown for $100\%$  ($50\%$) branching ratios.  ``j'' represents a jet that could be any light quark, for example $u, d, s, c$. Moreover, LHC luminosity for each search along with the experimental references are shown for each decay modes. Even though $LQ\overline{LQ}\to t\overline{t}\mu\overline{\mu}$ decay mode has the largest bound on LQ mass of 1420 GeV, within our scenario, the corresponding branching ratio is negligibly small, leading to a much lower mass bound. Consequently, the chosen leptoquark mass of $M_{LQ}= 1200$ GeV for our numerical analysis safely satisfies all collider bounds.

As for the diquark, LHC searches for dijets in the final state. Diquark mass smaller than 6 TeV is ruled out by recent  collider studies \cite{Khachatryan:2015dcf}. It should be pointed out that this limit was derived for diquarks that has couplings to up-quarks, which in our case only couples to down-quarks. Consequently, the lower bound on the mass is expected to be somewhat smaller. Not to mention, the limits largely depend on the associated branching ratios. In this work, for simplicity, we assume  its mass to be much  heavier compared to the  LQs such that all collider, as well as other experimental constraints are automatically satisfied. For example, among all processes mediated by the DQ,  the most dangerous constraint comes from its contribution to $B^0_s-\overline{B^0_s}$ mixing. Following  Ref. \cite{Bona:2007vi} we find very strong bounds on the Yukawa coupling that are given by:
\begin{align}
    \left| \left(y^{\omega}_{22} \right)^* y^{\omega}_{33} \right| < 2.177\times 10^{-3} \left( \frac{M_{\omega}}{\text{TeV}} \right)^2.
\end{align}
This for $M_{\omega}=10$ TeV leads to $\left| \left(y^{\omega}_{22} \right)^* y^{\omega}_{33} \right| < 0.2$. As can be easily verified from the benchmark points provided in Eqs. \eqref{BM-I} and \eqref{BM-II}, the types of solutions we achieve meet all requirements.

Before closing this section, we briefly discuss the possibility of LQ search at the future LHC within this setup.  As already aforementioned, at the LHC these LQs are produced mainly via the QCD driven pair-production, hence their production cross-sections only depend on their masses. Then at the 14 TeV LHC, the total  production cross-section is $\sigma (pp\to LQ \overline{LQ})= 2.2$ $fb$ for $M_{LQ}=1200$ GeV \cite{Kramer:2004df, Dumont:2016xpj}. Each of the pair-produced LQs will decay into a quark and a lepton, and we are interested in processes listed in Eqs. \eqref{br1}-\eqref{br4} for our example  \texttt{BM-II}. In the future, due to very high luminosity one would expect large number of  such events at the LHC that can potentially lead to the discovery of these leptoquarks.

For our benchmark values Eqs. \eqref{br1}-\eqref{br4}, the dominant decay modes of the LQs along with the corresponding branching ratios ($\beta$) are as follows:  
$S_1^{1/3}$ mostly decays to $s\nu$ ($\beta \simeq 29\%$) and $c\tau$ ($\beta \simeq 68\%$); $S^{4/3}_3$ to $b\tau$   ($\beta \simeq 85\%$);   $S^{-2/3}_3$ to $t\nu$   ($\beta\simeq 88\%$); $S^{1/3}_3$ mostly goes to $b\nu$   ($\beta\simeq 40\%$) and $t\tau$ ($\beta\simeq 45\%$). Then at the 3000 $fb^{-1}$ luminosity LHC run, one would except the following number of events: 
$\mathcal{N}\left[ pp\to S^{1/3}_1 ({S^{1/3}_1})^*\to (s\overline{s}) (\nu\overline{\nu})\right]=$ 
 555, $\mathcal{N}\left[ pp\to S^{1/3}_1 ({S^{1/3}_1})^*\to (c\overline{c}) (\tau\overline{\tau})\right]=$ 
 3050, and $\mathcal{N}\left[ pp\to S^{1/3}_1 ({S^{1/3}_1})^*\to (s\overline{c}) (\nu\overline{\tau})\right]=$ 
 1300; $\mathcal{N}\left[ pp\to S^{4/3}_3 ({S^{4/3}_3})^*\to (b\overline{b}) (\tau\overline{\tau})\right]=$ 
 4768; $\mathcal{N}\left[ pp\to S^{-2/3}_3 ({S^{-2/3}_3})^*\to (t\overline{t}) (\nu\overline{\nu})\right]=$ 
 5111;\\ $\mathcal{N}\left[ pp\to S^{1/3}_3 ({S^{1/3}_3})^*\to (b\overline{b}) (\nu\overline{\nu})\right]=$ 
 1056,  $\mathcal{N}\left[ pp\to S^{1/3}_3 ({S^{1/3}_3})^*\to (t\overline{t}) (\tau\overline{\tau})\right]=$ 
 1336, and $\mathcal{N}\left[ pp\to S^{1/3}_3 ({S^{1/3}_3})^*\to (b\overline{t}) (\nu\overline{\tau})\right]=$ 
 1188.

In the above, we have performed a simple estimation of the expected number of events at the high luminosity LHC run. Note however that the actual number of events that LHC will observe is somewhat less than that the numbers quoted above. This is due to several facts, for example, channels with bottom-quarks at the final state lose some efficiency due to b-tagging (for charm-quark, the c-tagging efficiency is even smaller).  This is also applicable for channels with top-quarks, since a top-quark will decay into a bottom-quark and jets (as well as charged leptons and neutrinos from the $W^{\pm}$ decay). Furthermore, one losses some efficiency in properly identifying missing energies associated to neutrinos in the final states. After taking all these effects into considerations, one needs to identify the corresponding SM backgrounds for comparison. However, it should be pointed out that the jets and the charged leptons produced directly from the LQ decay in the final states will be very hard due to heavy mass of the LQ. Hence they are expected to be detected in the LHC detectors more easily, as they can be separated from the jets and leptons produced from the SM processes, this also requires advanced cuts in event selections. Such a detailed collider study requires sophisticated simulation which is beyond the scope of this work, and we refer to the readers Refs. \cite{Blumlein:1996qp, Plehn:1997az, Kramer:1997hh, Eboli:1997fb, Kramer:2004df, Belyaev:2005ew,  Gripaios:2010hv, Davidson:2011zn, Hammett:2015sea, Mandal:2015vfa,  Mandal:2015lca, Evans:2015ita, Dumont:2016xpj, Khachatryan:2016jqo, Sirunyan:2017kqq, Dey:2017ede, Diaz:2017lit,Bandyopadhyay:2018syt,Sirunyan:2018nkj,Chandak:2019iwj} for dedicated search studies for leptoquarks.

\section{Conclusion}\label{SEC-06}
In this work we have explored the possibility that the neutrino mass, the long-standing tension in the  muon anomalous magnetic moment, and persistent observations of B-physics anomalies in the $R_{D^{(*)}}$, $R_{K^{(*)}}$ ratios have a common origin. Our proposal is a simple extension of the Standard Model that consists of two scalar leptoquarks $S_1\sim (\overline{3},1,1/3)$ and $S_3\sim (\overline{3},3,1/3)$, which are accompanied by a scalar diquark $\omega\sim (\overline{6},1,2/3)$. The muon receives a large
contribution towards its anomalous magnetic moment  due to chirality-enhanced effects from  leptoquark $S_1$ that explains $a_\mu$ data. This same   leptoquark $S_1$ also accommodates for the $R_{D^{(*)}}$ anomaly, whereas  leptoquark $S_3$ is responsible to account for the tension observed in the  $R_{K^{(*)}}$ ratio. Furthermore,  both $S_1$ and $S_3$ leptoquarks in association with the diquark $\omega$ participate in generating masses for the neutrinos at the two-loop order. A detailed analysis is carried out in this work, which shows strong correlations among various flavor violating processes, including neutrino oscillation parameters. In addition to exploring different regions in the parameter space of the theory, we have demonstrated the feasibility of this framework by providing benchmark points. These benchmark points successfully accommodate all three anomalies and naturally incorporate correct neutrino masses and mixings  while evading a number of experimental constraints from lepton flavor violation and flavor changing processes, as well as direct searches for leptoquakrs and diquarks at colliders. The lepton flavor violating rare decays of tau lepton $\tau\to \mu\gamma$, $\tau\to \mu\phi$, and $\tau\to \mu\eta$ are all predicted to be right below the current experimental upper bound. Other lepton flavor violating meson decays $B_s\to \tau \mu$ and $B\to K\tau\mu$ are expected to lie around one order below the present experimental limit as well. Hence, this model is very predictive and has the potential to be tested in near future by LHCb and Belle-II. Besides, the presence of TeV scale leptoquarks can lead the way to probe this model at the LHC in near future.

\section*{Acknowledgments}
We  thank Ahmed Ismail for useful discussion. 

\bibliographystyle{utphys}
\bibliography{references}
\end{document}